\documentclass[12pt]{article}

\usepackage{graphicx}
\usepackage{lscape}
\usepackage{citesort}

\begin{document}

\def\ds{\displaystyle}
\def\beq{\begin{equation}}
\def\eeq{\end{equation}}
\def\bea{\begin{eqnarray}}
\def\eea{\end{eqnarray}}
\def\beeq{\begin{eqnarray}}
\def\eeeq{\end{eqnarray}}
\def\ve{\vert}
\def\vel{\left|}
\def\ver{\right|}
\def\nnb{\nonumber}
\def\ga{\left(}
\def\dr{\right)}
\def\aga{\left\{}
\def\adr{\right\}}
\def\lla{\left<}
\def\rra{\right>}
\def\rar{\rightarrow}
\def\nnb{\nonumber}
\def\la{\langle}
\def\ra{\rangle}
\def\ba{\begin{array}}
\def\ea{\end{array}}
\def\tr{\mbox{Tr}}
\def\ssp{{\Sigma^{*+}}}
\def\sso{{\Sigma^{*0}}}
\def\ssm{{\Sigma^{*-}}}
\def\xis0{{\Xi^{*0}}}
\def\xism{{\Xi^{*-}}}
\def\qs{\la \bar s s \ra}
\def\qu{\la \bar u u \ra}
\def\qd{\la \bar d d \ra}
\def\qq{\la \bar q q \ra}
\def\gGgG{\la g^2 G^2 \ra}
\def\q{\gamma_5 \not\!q}
\def\x{\gamma_5 \not\!x}
\def\g5{\gamma_5}
\def\sb{S_Q^{cf}}
\def\sd{S_d^{be}}
\def\su{S_u^{ad}}
\def\ss{S_s^{??}}
\def\sbp{{S}_Q^{'cf}}
\def\sdp{{S}_d^{'be}}
\def\sup{{S}_u^{'ad}}
\def\ssp{{S}_s^{'??}}
\def\sig{\sigma_{\mu \nu} \gamma_5 p^\mu q^\nu}
\def\fo{f_0(\frac{s_0}{M^2})}
\def\ffi{f_1(\frac{s_0}{M^2})}
\def\fii{f_2(\frac{s_0}{M^2})}
\def\O{{\cal O}}
\def\sl{{\Sigma^0 \Lambda}}
\def\es{\!\!\! &=& \!\!\!}
\def\ap{\!\!\! &\approx& \!\!\!}
\def\ar{&+& \!\!\!}
\def\ek{&-& \!\!\!}
\def\kek{\!\!\!&-& \!\!\!}
\def\cp{&\times& \!\!\!}
\def\se{\!\!\! &\simeq& \!\!\!}
\def\eqv{&\equiv& \!\!\!}
\def\kpm{&\pm& \!\!\!}
\def\kmp{&\mp& \!\!\!}


\def\simlt{\stackrel{<}{{}_\sim}}
\def\simgt{\stackrel{>}{{}_\sim}}


\title{
         {\Large
                 {\bf
Polarized forward--backward asymmetries of leptons in 
$B_s \rar \ell^+ \ell^- \gamma$  decay 
                 }
         }
      }

\author{\vspace{1cm}\\
{\small T. M. Aliev \thanks
{e-mail: taliev@metu.edu}\,\,,
V. Bashiry \thanks
{e-mail: bashiry@newton.physics.metu.edu.tr}
\,\,,
M. Savc{\i} \thanks
{e-mail: savci@metu.edu}} \\
{\small Physics Department, Middle East Technical University,
06531 Ankara, Turkey} }

\date{}

\begin{titlepage}
\maketitle
\thispagestyle{empty}

\begin{abstract}
Polarized forward--backward asymmetries in the 
$B_s \rar \ell^+ \ell^- \gamma$  decay are calculated using the most 
general, model independent form of the effective Hamiltonian, including all
possible forms of interactions. The dependencies of the asymmetries on new
Wilson coefficients are investigated. The detectability of the asymmetries
at LHC is discussed.
\end{abstract}
~~~PACS numbers: 12.60.--i, 13.30.--a
\end{titlepage}

\section{Introduction}

Rare radiative leptonic $B_{s(d)} \rar \ell^+ \ell^- \gamma$  decays are 
induced by the flavor--changing neutral current transitions $b\rar s(d)$. 
In the standard model (SM) such processes are described by the penguin and 
box diagrams and have branching ratios $10^{-8}-10^{-15}$ (see for example
\cite{R6701}). These rare decays can not be observed at the running machines
such as Tevatron, BaBar and Belle, but the $B_{s(d)} \rar \mu^+ \mu^-$ and 
$B_{s(d)} \rar \mu^+ \mu^- \gamma$  decays can be detected at LHC with ATLAS,
CMS and LHCb detectors \cite{R6702}. 
Many experimental observables such
as, the branching ratio, photon energy, dilepton mass spectra and charge
asymmetries, as well as the transition form factors, are investigated for 
the $B_{s(d)} \rar \ell^+ \ell^- \gamma$ 
decays in \cite{R6703,R6704,R6705,R6706,R6707,R6708,R6709}. 
At the same time $B_{s(d)} \rar \ell^+ \ell^-
\gamma$  decays might be sensitive to the new physics beyond the SM. New
physics effects in these decays can appear in two different ways: either
through the new operators in the effective Hamiltonian which are absent in
the SM, or through new contributions to the Wilson coefficients existing in
the SM. One efficient way for precise determination of the SM parameters and
looking for new physics beyond the SM is studying the lepton polarization
effects. It has been pointed out in \cite{R6710} that some of the single
lepton polarization asymmetries might be too small to be observed and might
not provide sufficient number of observables in checking the structure of
the effective Hamiltonian. In need of more observables, in \cite{R6710},
the maximum number of independent observables have been constructed 
by considering the situation where both lepton polarizations are 
simultaneously measured.

In the present work, we analyze the possibility of searching for new physics
in the $B_s \rar \ell^+ \ell^- \gamma$  decay by studying the forward--backward
asymmetries when both leptons are polarized, using the most general, model 
independent form of the effective Hamiltonian including all possible interactions.  
Note that the sensitivity of double--lepton polarization asymmetries on new
Wilson coefficients for the $B_s \rar \ell^+ \ell^- \gamma$  decay has been
investigated recently in \cite{R6711}. 

The work is organized as follows. In section 2, the matrix element for the
$B_s \rar \ell^+ \ell^- \gamma$  is obtained, using the general, model
independent form of the effective Hamiltonian. In section 3, we calculate
the polarized forward--backward asymmetries of the leptons in $B_s \rar \ell^+
\ell^- \gamma$  decay. Section for is devoted to the
numerical analysis, discussions and conclusions.

\section{Theoretical framework}

In the present section we derive the matrix element for the $B_s \rar \ell^+ \ell^-
\gamma$  using the general, model independent form of the effective
Hamiltonian. The matrix element for the process $B_s \rar \ell^+ \ell^- \gamma$     
can be obtained from that of the purely leptonic $B_s \rar \ell^+ \ell^-$  decay. 
At inclusive level the process $B_s \rar \ell^+ \ell^-$ is described by 
$b \rar q \ell^+ \ell^-$ transition. The effective $b\rar q \ell^+ \ell^-$ 
transition can be written in terms of twelve model independent
four--Fermi interactions in the following form \cite{R6712}:
\bea
\label{e6701}
{\cal H}_{eff} \es \frac{G\alpha}{\sqrt{2} \pi}
 V_{tq}V_{tb}^\ast
\Bigg\{ C_{SL} \, \bar q i \sigma_{\mu\nu} \frac{q^\nu}{q^2}\, L \,b
\, \bar \ell \gamma^\mu \ell + C_{BR}\, \bar q i \sigma_{\mu\nu}
\frac{q^\nu}{q^2} \,R\, b \, \bar \ell \gamma^\mu \ell \nnb \\
\ar C_{LL}^{tot}\, \bar q \gamma_\mu L b \,\bar \ell \gamma^\mu L \ell +
C_{LR}^{tot} \,\bar q \gamma_\mu L b \, \bar \ell \gamma^\mu R \ell +
C_{RL} \,\bar q \gamma_\mu R b \,\bar \ell \gamma^\mu L \ell \nnb \\
\ar C_{RR} \,\bar q \gamma_\mu R b \, \bar \ell \gamma^\mu R \ell +
C_{LRLR} \, \bar q R b \,\bar \ell R \ell +
C_{RLLR} \,\bar q L b \,\bar \ell R \ell
+ C_{LRRL} \,\bar q R b \,\bar \ell L \ell \nnb \\
\ar C_{RLRL} \,\bar q L b \,\bar \ell L \ell+
C_T\, \bar q \sigma_{\mu\nu} b \,\bar \ell \sigma^{\mu\nu}\ell 
+ i C_{TE}\,\epsilon^{\mu\nu\alpha\beta} \bar q \sigma_{\mu\nu} b \,
\bar \ell \sigma_{\alpha\beta} \ell  \Bigg\}~,
\eea
where $C_X$ are the coefficients of the four--Fermi interactions and
\bea
L = \frac{1-\gamma_5}{2} ~,~~~~~~ R = \frac{1+\gamma_5}{2}\nnb~.
\eea
The terms with coefficients $C_{SL}$ and $C_{BR}$ which describe penguin
contributions correspond to $-2 m_s C_7^{eff}$ 
and $-2 m_b C_7^{eff}$ in the SM, respectively. The next four terms in this
expression are the vector interactions. The interaction terms containing
$C_{LL}^{tot}$ and $C_{LR}^{tot}$ in the SM have the form
$C_9^{eff}-C_{10}$ and $C_9^{eff}+C_{10}$, respectively. Inspired by this 
$C_{LL}^{tot}$ and $C_{LR}^{tot}$ will be written as
\bea
C_{LL}^{tot} \es C_9^{eff} - C_{10} + C_{LL}~, \nnb \\
C_{LR}^{tot} \es C_9^{eff} + C_{10} + C_{LR}~, \nnb
\eea
where $C_{LL}$ and $C_{LR}$ describe contributions from new physics.
The terms with coefficients $C_{LRLR}$, $C_{RLLR}$, $C_{LRRL}$ and 
$C_{RLRL}$ describe the scalar type interactions. The last two 
terms in Eq. (\ref{e6701}) with the coefficients $C_T$ and $C_{TE}$
describe the tensor type interactions.

Having presented the general form of the effective Hamiltonian the next
problem is the calculation of the matrix element of the $B_q \rar \ell^+ \ell^-
\gamma$  decay. This matrix element can be written as the sum of the two
parts, structure--dependent and inner--Bremsstrahlung parts 
\bea
\label{e6702}
{\cal M} = {\cal M}_{SD}+{\cal M}_{IB}~. 
\eea
The matrix element for the structure--dependent part ${\cal M}_{SD}$, which
corresponds to the radiation of photon from initial quarks, can be obtained
by calculating the matrix element $\lla \gamma\vel{\cal H}_{eff}\ver B \rra$.
Using Eq. (\ref{e6701}) we see that, for calculation of 
${\cal M}_{SD}$, we need to know the following matrix elements
\bea
\label{e6703}
&&\lla \gamma\vel \bar s \gamma_\mu (1 \mp \gamma_5)
b \ver B \rra~,\nnb \\
&&\lla \gamma \vel \bar s \sigma_{\mu\nu} q^\nu b \ver B \rra~, \nnb \\
&&\lla \gamma \vel \bar s \sigma_{\mu\nu} b \ver B \rra~, \nnb \\
&&\lla \gamma \vel \bar s (1 \mp \gamma_5) b
\ver B \rra~.
\eea
The first two of the matrix elements in Eq. (\ref{e6703}) are defined in the
following way \cite{R6703,R6707,R6713,R6714}
\bea
\label{e6704}
\lla \gamma(k) \vel \bar q \gamma_\mu
(1 \mp \gamma_5) b \ver B(p_B) \rra \es
\frac{e}{m_B^2} \Big\{
\epsilon_{\mu\nu\lambda\sigma} \varepsilon^{\ast\nu} q^\lambda
k^\sigma g(q^2) \pm i\,
\Big[ \varepsilon^{\ast\mu} (k q) -
(\varepsilon^\ast q) k^\mu \Big] f(q^2) \Big\}~,\\ \nnb \\
\label{e6705}
\lla \gamma(k) \vel \bar q \sigma_{\mu\nu} b \ver B(p_B) \rra \es 
\frac{e}{m_B^2}
\epsilon_{\mu\nu\lambda\sigma} \Big[
G \varepsilon^{\ast\lambda} k^\sigma +
H \varepsilon^{\ast\lambda} q^\sigma +
N (\varepsilon^\ast q) q^\lambda k^\sigma \Big]~.
\eea
Here, $\varepsilon^\ast$ and $k$ are the four vector polarization
and momentum of the photon, respectively, $q=p_B-k$ is the momentum transfer,
$p_B$ is the momentum of the $B$ meson and $g(q^2)$, $f(q^2)$, $G(q^2)$, 
$H(q^2)$ and $N(q^2)$ are the $B_s \rar \gamma$  transition form factors. 
The matrix element 
$\lla \gamma(k) \vel \bar s \sigma_{\mu\nu} \gamma_5 b \ver B(p_B) \rra$
can be obtained from Eq. (\ref{e6705}) using the identity
\bea
\sigma_{\mu\nu} = - \frac{i}{2}\epsilon_{\mu\nu\alpha\beta}
\sigma^{\alpha\beta} \gamma_5~.\nnb
\eea
The matrix elements 
$\lla \gamma(k) \vel \bar s (1 \mp \gamma_5) b \ver B(p_B) \rra$ 
and
$\lla \gamma \vel \bar s i \sigma_{\mu\nu} q^\nu b \ver B \rra$ can be
obtained from Eqs. (\ref{e6704}) and (\ref{e6705})
by multiplying them $q^\mu$ and $q^\nu$, respectively, as a result of which
we get
\bea
\label{e6706}
\lla \gamma(k) \vel \bar s (1 \mp \gamma_5) b \ver B(p_B) \rra
\es 0~, \\
\label{e6707}
\lla \gamma \vel \bar s i \sigma_{\mu\nu} q^\nu b \ver B \rra \es
\frac{e}{m_B^2} i\, \epsilon_{\mu\nu\alpha\beta} q^\nu
\varepsilon^{\alpha\ast} k^\beta G~.
\eea
The matrix element $\lla \gamma
\vel \bar s i \sigma_{\mu\nu} q^\nu (1+\gamma_5) b \ver B \rra$ can be
written in terms of the two form factors $f_1(q^2)$ and $g_1(q^2)$
that are calculated in the framework of QCD
sum rules \cite{R6703,R6713} in the following way 
\bea
\label{e6708}
\lla \gamma \vel \bar s i \sigma_{\mu\nu} q^\nu (1+\gamma_5) b \ver B \rra \es
\frac{e}{m_B^2} \Big\{
\epsilon_{\mu\alpha\beta\sigma} \, \varepsilon^{\alpha\ast} q^\beta k^\sigma
g_1(q^2)
+ i\,\Big[\varepsilon_\mu^\ast (q k) - (\varepsilon^\ast q) k_\mu \Big]
f_1(q^2) \Big\}~.
\eea
It should be noted that these form factors were calculated in framework of
the light--front model in \cite{R6714}.
Eqs. (\ref{e6705}), (\ref{e6707}) and (\ref{e6708}) allow us to
express $G,~H$ and $N$ in terms of $f_1$ and $g_1$. 
Eqs. (\ref{e6704})--(\ref{e6708}) help us rewrite ${\cal M}_{SD} $ in the
following form
\bea
\label{e6709}
{\cal M}_{SD} \es \frac{\alpha G_F}{4 \sqrt{2} \, \pi} V_{tb} V_{tq}^* 
\frac{e}{m_B^2} \,\Bigg\{
\bar \ell \gamma^\mu (1-\gamma_5) \ell \, \Big[
A_1 \epsilon_{\mu \nu \alpha \beta} 
\varepsilon^{\ast\nu} q^\alpha k^\beta + 
i \, A_2 \Big( \varepsilon_\mu^\ast (k q) - 
(\varepsilon^\ast q ) k_\mu \Big) \Big] \nnb \\
\ar \bar \ell \gamma^\mu (1+\gamma_5) \ell \, \Big[
B_1 \epsilon_{\mu \nu \alpha \beta} 
\varepsilon^{\ast\nu} q^\alpha k^\beta 
+ i \, B_2 \Big( \varepsilon_\mu^\ast (k q) - 
(\varepsilon^\ast q ) k_\mu \Big) \Big] \nnb \\
\ar i \, \epsilon_{\mu \nu \alpha \beta} 
\bar \ell \sigma^{\mu\nu}\ell \, \Big[ G \varepsilon^{\ast\alpha} k^\beta 
+ H \varepsilon^{\ast\alpha} q^\beta + 
N (\varepsilon^\ast q) q^\alpha k^\beta \Big] \\
\ar i \,\bar \ell \sigma_{\mu\nu}\ell \, \Big[
G_1 (\varepsilon^{\ast\mu} k^\nu - \varepsilon^{\ast\nu} k^\mu) + 
H_1 (\varepsilon^{\ast\mu} q^\nu - \varepsilon^{\ast\nu} q^\mu) +
N_1 (\varepsilon^\ast q) (q^\mu k^\nu - q^\nu k^\mu) \Big] \Bigg\}~,\nnb
\eea
where
\bea
\label{e6710}
A_1 \es \frac{1}{q^2} \Big( C_{BR} + C_{SL} \Big) g_1 +
\Big( C_{LL}^{tot} + C_{RL} \Big) g ~, \nnb \\
A_2 \es \frac{1}{q^2} \Big( C_{BR} - C_{SL} \Big) f_1 +
\Big( C_{LL}^{tot} - C_{RL} \Big) f ~, \nnb \\
B_1 \es \frac{1}{q^2} \Big( C_{BR} + C_{SL} \Big) g_1 +
\Big( C_{LR}^{tot} + C_{RR} \Big) g ~, \nnb \\
B_2 \es \frac{1}{q^2} \Big( C_{BR} - C_{SL} \Big) f_1 +
\Big( C_{LR}^{tot} - C_{RR} \Big) f ~, \nnb \\
G \es 4 C_T g_1 ~, \nnb \\
N \es - 4 C_T \frac{1}{q^2} (f_1+g_1) ~, \\
H \es N (qk) ~, \nnb \\
G_1 \es - 8 C_{TE} g_1 ~, \nnb \\
N_1 \es 8 C_{TE} \frac{1}{q^2} (f_1+g_1) ~, \nnb \\ 
H_1 \es N_1(qk)~. \nnb
\eea
In regard to the inner--Bremsstrahlung part, as a result of relevant
calculations we get
\bea
\label{e6711}
{\cal M}_{IB} \es \frac{\alpha G_F}{4 \sqrt{2} \, \pi} V_{tb} V_{tq}^*  
e f_B i \,\Bigg\{
F\, \bar \ell  \Bigg(
\frac{{\not\!\varepsilon}^\ast {\not\!p}_B}{2 p_1 k} - 
\frac{{\not\!p}_B {\not\!\varepsilon}^\ast}{2 p_2 k} \Bigg) 
\gamma_5 \ell \nnb \\
\ar F_1 \, \bar \ell  \Bigg[
\frac{{\not\!\varepsilon}^\ast {\not\!p}_B}{2 p_1 k} -
\frac{{\not\!p}_B {\not\!\varepsilon}^\ast}{2 p_2 k} +
2 m_\ell \Bigg(\frac{1}{2 p_1 k} + \frac{1}{2 p_2 k}\Bigg)
{\not\!\varepsilon}^\ast \Bigg] \ell \Bigg\}~.
\eea
In deriving Eq. (\ref{e6711}), we have used
\bea
\la 0 \ve \bar s \gamma_\mu \gamma_5 b \ve B \ra \es 
-~i f_B p_{B\mu}~, \nnb \\
\la 0 \ve \bar s \sigma_{\mu\nu} (1+\gamma_5) b \ve B \ra \es 0~,\nnb
\eea
The functions $F$ and $F_1$ are defined as follows
\bea
\label{e6712}
F \es 2 m_\ell \Big( C_{LR}^{tot} - C_{LL}^{tot} + C_{RL} - C_{RR} \Big)
+ \frac{m_B^2}{m_b}
\Big( C_{LRLR} - C_{RLLR} - C_{LRRL} + C_{RLRL} \Big)~, \nnb \\
F_1 \es\frac{m_B^2}{m_b} \Big( C_{LRLR} - C_{RLLR} + C_{LRRL} - C_{RLRL}
\Big)~.
\eea

\section{Polarized forward--backward asymmetries of the leptons in
$B_s \rar \ell^+ \ell^- \gamma$  decay}

In the present section we calculate the polarized forward--backward 
asymmetries of leptons. For this purpose we
define the following orthogonal unit vectors $s_i^\pm$ (here $i = L,~T$ or
$N$ stands for longitudinal, transversal or normal polarizations,
respectively) in the rest frame of $\ell^\pm$
\bea
\label{e6713}
s^{-\mu}_L \es \ga 0,\vec{e}_L^{\,-}\dr =
\ga 0,\frac{\vec{p}_-}{\vel\vec{p}_- \ver}\dr~, \nnb \\
s^{-\mu}_N \es \ga 0,\vec{e}_N^{\,-}\dr = \ga 0,\frac{\vec{p}_\Lambda\times
\vec{p}_-}{\vel \vec{p}_\Lambda\times \vec{p}_- \ver}\dr~, \nnb \\
s^{-\mu}_T \es \ga 0,\vec{e}_T^{\,-}\dr = \ga 0,\vec{e}_N^{\,-}
\times \vec{e}_L^{\,-} \dr~, \nnb \\
s^{+\mu}_L \es \ga 0,\vec{e}_L^{\,+}\dr =
\ga 0,\frac{\vec{p}_+}{\vel\vec{p}_+ \ver}\dr~, \nnb \\ 
s^{+\mu}_N \es \ga 0,\vec{e}_N^{\,+}\dr = \ga 0,\frac{\vec{p}_\Lambda\times
\vec{p}_+}{\vel \vec{p}_\Lambda\times \vec{p}_+ \ver}\dr~, \nnb \\
s^{+\mu}_T \es \ga 0,\vec{e}_T^{\,+}\dr = \ga 0,\vec{e}_N^{\,+}
\times \vec{e}_L^{\,+}\dr~,
\eea
where $\vec{p}_\pm$ and $\vec{k}$ are the three--momenta of the
leptons $\ell^\pm$ and photon in the
center of mass frame (CM) of $\ell^- \,\ell^+$ system, respectively.
Transformation of unit vectors from the rest frame of the leptons to CM
frame of leptons can be accomplished by the Lorentz boost. Boosting of the
longitudinal unit vectors $s_L^{\pm\mu}$ yields
\bea
\label{e6714}
\ga s^{\mp\mu}_L \dr_{CM} \es \ga \frac{\vel\vec{p}_\mp \ver}{m_\ell}~,
\frac{E_\ell \vec{p}_\mp}{m_\ell \vel\vec{p}_\mp \ver}\dr~,
\eea
where $\vec{p}_+ = - \vec{p}_-$, $E_\ell$ and $m_\ell$ are the energy
mass of leptons in the CM frame.
The remaining unit vectors $s_N^{\pm\mu}$, $s_T^{\pm\mu}$ are unchanged
under Lorentz transformation.

The definition of the normalized, unpolarized differential
forward--backward asymmetry is
\bea
\label{e6715}
{\cal A}_{FB} = \frac{\ds \int_{0}^{1} \frac{d^2\Gamma}{d\hat{s} dz} -
\int_{-1}^{0} \frac{d^2\Gamma}{d\hat{s} dz}}
{\ds \int_{0}^{1} \frac{d^2\Gamma}{d\hat{s} dz} +
\int_{-1}^{0} \frac{d^2\Gamma}{d\hat{s} dz}}~,
\eea
where $z=\cos\theta$ is the angle between $\Lambda_b$ meson and $\ell^-$ in the
center mass frame of leptons. When the spins of both leptons are taken into
account, the ${\cal A}_{FB}$ will be a function of the spins of the final
leptons and it is defined as
\bea
\label{e6716}
{\cal A}_{FB}^{ij}(\hat{s}) \es 
\Bigg(\frac{d\Gamma(\hat{s})}{d\hat{s}} \Bigg)^{-1}
\Bigg\{ \int_0^1 dz - \int_{-1}^0 dz \Bigg\}
\Bigg\{ 
\Bigg[
\frac{d^2\Gamma(\hat{s},\vec{s}^{\,-} = \vec{i},\vec{s}^{\,+} = \vec{j})}
{d\hat{s} dz} - 
\frac{d^2\Gamma(\hat{s},\vec{s}^{\,-} = \vec{i},\vec{s}^{\,+} = -\vec{j})} 
{d\hat{s} dz}
\Bigg] \nnb \\
\ek
\Bigg[
\frac{d^2\Gamma(\hat{s},\vec{s}^{\,-} = -\vec{i},\vec{s}^{\,+} = \vec{j})} 
{d\hat{s} dz} - 
\frac{d^2\Gamma(\hat{s},\vec{s}^{\,-} = -\vec{i},\vec{s}^{\,+} = -\vec{j})} 
{d\hat{s} dz}
\Bigg]
\Bigg\}~,\nnb \\ \nnb \\
\es 
{\cal A}_{FB}(\vec{s}^{\,-}=\vec{i},\vec{s}^{\,+}=\vec{j})   -
{\cal A}_{FB}(\vec{s}^{\,-}=\vec{i},\vec{s}^{\,+}=-\vec{j})  - 
{\cal A}_{FB}(\vec{s}^{\,-}=-\vec{i},\vec{s}^{\,+}=\vec{j})  \nnb \\
\ar   
{\cal A}_{FB}(\vec{s}^{\,-}=-\vec{i},\vec{s}^{\,+}=-\vec{j})~.   
\eea

Using these definitions for the double polarized $FB$ asymmetries, we get
the following results:   

\bea
\label{e6717}
{\cal A}_{FB}^{LL} \es 
\frac{1}{\Delta} \Bigg\{
- 4 m_B^2 \hat{s} (1-\hat{s})^2 v \mbox{\rm Re} 
[A_1^\ast A_2 - B_1^\ast B_2] \nnb \\
\ek  \frac{2}{\hat{m}_\ell} m_B \hat{s} (1-\hat{s})^2 v (1-v^2)
\Big( \mbox{\rm Im}[(A_1^\ast-B_1^\ast) G_1] - \mbox{\rm Re}[(A_2-B_2)^\ast G]
\Big) \nnb \\
\ek \frac{4}{\hat{m}_\ell} m_B \hat{s}^2 (1-\hat{s}) v (1-v^2)
\mbox{\rm Im}[(A_1^\ast-B_1^\ast) H_1] \nnb \\
\ar \frac{4}{\hat{m}_\ell v} f_B m_B \hat{s} (1-\hat{s}) (1-v^2)
\ln [1-v^2] \mbox{\rm Re}[(A_2^\ast-B_2^\ast) F] \nnb \\  
\ek  \frac{4}{\hat{m}_\ell v} f_B m_B \hat{s} (1-\hat{s}) (1-v^2)
\ln [1-v^2] \mbox{\rm Re}[(A_1^\ast-B_1^\ast) F_1] \Bigg\}~, \\ \nnb \\
\label{e6718}
{\cal A}_{FB}^{LN} \es
\frac{1}{\Delta} \Bigg\{
-\frac{4}{3} m_B \sqrt{\hat{s}}(1-\hat{s})^2 v \mbox{\rm Re}[
(A_1^\ast- A_2^\ast + B_1^\ast + B_2^\ast ) G_1] \nnb \\
\ar \frac{4}{3} m_B \sqrt{\hat{s}}(1-\hat{s})^2 v \mbox{\rm Im}[
(A_1^\ast- A_2^\ast - B_1^\ast - B_2^\ast ) G] \nnb \\
\ar \frac{4}{3} m_B^3 \sqrt{\hat{s}^3}(1-\hat{s})^2 v \Big(\mbox{\rm Re}[
(A_2^\ast - B_2^\ast ) N_1^\ast] 
- \mbox{\rm Im}[(A_2^\ast + B_2^\ast ) N] \Big) \nnb \\
\ek \frac{2}{3 \hat{m}_\ell} m_B^2 \sqrt{\hat{s}^3}(1-\hat{s})^2 v (1-v^2)
\Big( 2 \mbox{\rm Re}[G^\ast N_1 + G_1^\ast N + m_B^2 \hat{s} N_1^\ast N]  
+\mbox{\rm Im}[A_1^\ast B_1 + A_2^\ast B_2] \Big) \nnb \\
\ek f_B m_B^2 \sqrt{\hat{s}} (1-\hat{s})
\Big\{ 2 f_B m_B^2 \hat{m}_\ell    
\mbox{\rm Im}[F_1^\ast F] I_4 +
v \Big[ m_B (1-\hat{s}) \mbox{\rm Im}[(A_1^\ast +B_1^\ast ) F_1] \nnb \\
\ar m_B \mbox{\rm Im}[(A_1^\ast-A_2^\ast-B_1^\ast-B_2^\ast) F - \hat{s}       
(A_1^\ast+A_2^\ast-B_1^\ast+B_2^\ast) F]
+ 8 \hat{m}_\ell \mbox{\rm Re}[F^\ast H_1] \Big] I_7 \Big\} \nnb \\
\ar f_B m_B^3 \sqrt{\hat{s}} (1-\hat{s}) v [1- \hat{s} (1-2 v^2)] 
\mbox{\rm Im}[(A_2^\ast - B_2^\ast) F_1] {\cal J}_{4} \nnb \\
\ar 8 f_B m_B^2 \hat{m}_\ell \sqrt{\hat{s}} (1-\hat{s}) v 
\mbox{\rm Re}[F^\ast (G_1 + m_B^2 N_1)] {\cal J}_{4} \nnb \\
\ar 4 f_B m_B^4 \hat{m}_\ell \sqrt{\hat{s}} (1-\hat{s})^2 v
\mbox{\rm Im}[F_1^\ast N] {\cal J}_{4} \Bigg\}~, \\ \nnb \\
\label{e6719}
{\cal A}_{FB}^{NL} \es
\frac{1}{\Delta} \Bigg\{
\frac{4}{3} m_B \sqrt{\hat{s}}(1-\hat{s})^2 v \mbox{\rm Re}[
(A_1^\ast+ A_2^\ast + B_1^\ast - B_2^\ast ) G_1] \nnb \\
\ar \frac{4}{3} m_B \sqrt{\hat{s}}(1-\hat{s})^2 v \mbox{\rm Im}[
(A_1^\ast+ A_2^\ast - B_1^\ast + B_2^\ast ) G] \nnb\\
\ar \frac{4}{3} m_B^3 \sqrt{\hat{s}^3}(1-\hat{s})^2 v \Big(\mbox{\rm Re}[
(A_2^\ast - B_2^\ast ) N_1] 
+ \mbox{\rm Im}[(A_2^\ast + B_2^\ast ) N] \Big) \nnb \\
\ar \frac{2}{3 \hat{m}_\ell} m_B^2 \sqrt{\hat{s}^3}(1-\hat{s})^2 v (1-v^2) 
\Big( 2 \mbox{\rm Re}[G^\ast N_1 + G_1^\ast N + m_B^2 \hat{s} N_1^\ast N]
-\mbox{\rm Im}[A_1^\ast B_1 + A_2^\ast B_2] \Big) \nnb \\
\ar f_B m_B^2 \sqrt{\hat{s}} (1-\hat{s}) 
\Big\{ 2 f_B m_B^2 \hat{m}_\ell    
\mbox{\rm Im}[F_1^\ast F] I_4 +
v \Big[m_B (1-\hat{s}) \mbox{\rm Im}[(A_1^\ast +B_1^\ast ) F_1] \nnb \\
\ek m_B \mbox{\rm Im}[(A_1^\ast+A_2^\ast-B_1^\ast+B_2^\ast) F - \hat{s}       
(A_1^\ast-A_2^\ast-B_1^\ast-B_2^\ast) F]
+ 8 \hat{m}_\ell \mbox{\rm Re}[F^\ast H_1] \Big] I_7 \Big\} \nnb \\
\ar f_B m_B^3 \sqrt{\hat{s}} (1-\hat{s}) v [1- \hat{s} (1-2 v^2)] 
\mbox{\rm Im}[(A_2^\ast - B_2^\ast) F_1] {\cal J}_{4}\nnb \\
\ek 8 f_B m_B^2 \hat{m}_\ell \sqrt{\hat{s}} (1-\hat{s}) v
\mbox{\rm Re}[F^\ast (G_1 + m_B^2 N_1)] {\cal J}_{4} \nnb \\
\ek 4 f_B m_B^4 \hat{m}_\ell \sqrt{\hat{s}} (1-\hat{s})^2 v
\mbox{\rm Im}[F_1^\ast N] {\cal J}_{4} \Bigg\}~, \\ \nnb \\
\label{e6720}                                                                
{\cal A}_{FB}^{LT} \es
\frac{1}{\Delta} \Bigg\{
\frac{4}{3 \sqrt{\hat{s}}} \hat{m}_\ell (1-\hat{s})^2 
\Big[ 4 \ga \vel G_1\ver^2 + \vel G \ver^2\dr + m_B^2 \hat{s}
\ga \vel A_1 \ver^2 + \vel A_2 \ver^2 + \vel B_1 \ver^2 +
\vel B_2 \ver^2 \dr \Big] \nnb \\
\ek \frac{4}{3} m_B \sqrt{\hat{s}} (1-\hat{s})^2 v^2 
\Big( \mbox{\rm Im}[(A_1^\ast-B_1^\ast) G_1] - 
\mbox{\rm Re}[(A_2^\ast-B_2^\ast) (G + m_B^2 \hat{s} N)] \Big) \nnb \\
\ek \frac{4}{3} m_B \sqrt{\hat{s}} (1-\hat{s})^2 (2-v^2)
\Big( \mbox{\rm Re}[(A_1^\ast+B_1^\ast) G]
- \mbox{\rm Im}[(A_2^\ast+B_2^\ast) (G_1 + m_B^2 \hat{s} N_1] \Big) \nnb \\
\ar
\frac{8}{3} m_B^2 \hat{m}_\ell \sqrt{\hat{s}} (1-\hat{s})^2        
\Big( \mbox{\rm Re}[A_1^\ast B_1 + A_2^\ast B_2 + 4 G_1^\ast N_1]
+ 2 m_B^2 \hat{s} \vel N_1 \ver^2 \Big) \nnb \\
\ar \frac{1}{\sqrt{\hat{s}}} f_B^2 m_B^4 \hat{m}_\ell (1-\hat{s})
\Big[ (1-\hat{s}) \ga \vel F_1 \ver^2 + \vel F \ver^2 \dr 
({\cal J}_{1} + {\cal J}_{2}) + 2 \hat{s} v \vel F_1 \ver^2 {\cal J}_{3} \Big]\nnb \\
\ar f_B m_B^3 \sqrt{\hat{s}} (1-\hat{s})^2 v^2 \mbox{\rm Re}[(A_1^\ast-B_1^\ast) F_1] 
{\cal J}_{4} \nnb \\
\ek f_B m_B^3 \sqrt{\hat{s}} (1-\hat{s}^2) v^2 
\mbox{\rm Re}[(A_2^\ast-B_2^\ast) F^\ast]
{\cal J}_{4} \nnb \\
\ar f_B m_B^3 \sqrt{\hat{s}} (1-\hat{s})^2 (2-v^2) 
\mbox{\rm Re}[(A_1^\ast+B_1^\ast) F] {\cal J}_{4} \nnb \\
\ek 4 f_B m_B^4 \hat{m}_\ell \sqrt{\hat{s}} (1-\hat{s})
[2 + v^2 - \hat{s} (2 - v^2)]
\mbox{\rm Im}[F_1^\ast N_1] {\cal J}_{4} \nnb \\
\ar 8 f_B m_B^2 \hat{m}_\ell \sqrt{\hat{s}} (1-\hat{s}) v^2 
\mbox{\rm Im}[F_1^\ast H_1] {\cal J}_{4} \nnb \\
\ek f_B m_B^3 \sqrt{\hat{s}} (1-\hat{s}) [2 - v^2 - \hat{s} (2-3 v^2)]
\mbox{\rm Re}[(A_2^\ast+B_2^\ast) F_1] {\cal J}_{4} \nnb \\
\ek \frac{8}{\sqrt{\hat{s}}} f_B m_B^2 \hat{m}_\ell (1-\hat{s})
\Big[ (1-\hat{s} + \hat{s} v^2) \mbox{\rm Im}[F_1^\ast G_1] +
(1-\hat{s}) \mbox{\rm Re}[F^\ast G] \Big] {\cal J}_{4} \Bigg\}~, \\ \nnb \\
\label{e6721}
{\cal A}_{FB}^{TL} \es
\frac{1}{\Delta} \Bigg\{
- \frac{4}{3 \sqrt{\hat{s}}} \hat{m}_\ell (1-\hat{s})^2 
\Big[ 4 \ga \vel G_1\ver^2 + \vel G \ver^2\dr + m_B^2 \hat{s}
\ga \vel A_1 \ver^2 + \vel A_2 \ver^2 + \vel B_1 \ver^2 +
\vel B_2 \ver^2 \dr \Big] \nnb \\
\ek \frac{4}{3} m_B \sqrt{\hat{s}} (1-\hat{s})^2 v^2 
\Big( \mbox{\rm Im}[(A_1^\ast-B_1^\ast) G_1] - 
\mbox{\rm Re}[(A_2^\ast-B_2^\ast) (G + m_B^2 \hat{s} N)] \Big) \nnb \\
\ar \frac{4}{3} m_B \sqrt{\hat{s}} (1-\hat{s})^2 (2-v^2)
\Big( \mbox{\rm Re}[(A_1^\ast+B_1^\ast) G]
- \mbox{\rm Im}[(A_2^\ast+B_2^\ast) (G_1 + m_B^2 \hat{s} N_1] \Big) \nnb \\
\ek \frac{8}{3} m_B^2 \hat{m}_\ell \sqrt{\hat{s}} (1-\hat{s})^2
\Big( \mbox{\rm Re}[A_1^\ast B_1 + A_2^\ast B_2 + 4 G_1^\ast N_1]
+ 2 m_B^2 \hat{s} \vel N_1 \ver^2 \Big) \nnb \\
\ek \frac{1}{\sqrt{\hat{s}}} f_B^2 m_B^4 \hat{m}_\ell (1-\hat{s})
\Big[ (1-\hat{s}) \ga \vel F_1 \ver^2 + \vel F \ver^2 \dr 
({\cal J}_{1} + {\cal J}_{2}) + 2 \hat{s} v \vel F_1 \ver^2 {\cal J}_{3} \Big] \nnb \\
\ar f_B m_B^3 \sqrt{\hat{s}} (1-\hat{s})^2 v^2 
\mbox{\rm Re}[(A_1^\ast-B_1^\ast) F_1] {\cal J}_{4} \nnb \\
\ek f_B m_B^3 \sqrt{\hat{s}} (1-\hat{s}^2) v^2 
\mbox{\rm Re}[(A_2^\ast-B_2^\ast) F]
{\cal J}_{4} \nnb \\
\ek f_B m_B^3 \sqrt{\hat{s}} (1-\hat{s})^2 (2-v^2) 
\mbox{\rm Re}[(A_1^\ast+B_1^\ast) F] {\cal J}_{4} \nnb \\
\ar 4 f_B m_B^4 \hat{m}_\ell \sqrt{\hat{s}} (1-\hat{s})
[2 + v^2 - \hat{s} (2 - v^2)]   
\mbox{\rm Im}[F_1^\ast N_1] {\cal J}_{4} \nnb \\
\ek 8 f_B m_B^2 \hat{m}_\ell \sqrt{\hat{s}} (1-\hat{s}) v^2 
\mbox{\rm Im}[F_1^\ast H_1] {\cal J}_{4} \nnb \\
\ar f_B m_B^3 \sqrt{\hat{s}} (1-\hat{s}) [2 - v^2 - \hat{s} (2-3 v^2)]
\mbox{\rm Re}[(A_2^\ast+B_2^\ast) F_1] {\cal J}_{4} \nnb \\
\ar \frac{8}{\sqrt{\hat{s}}} f_B m_B^2 \hat{m}_\ell (1-\hat{s})
\Big[ (1-\hat{s} + \hat{s} v^2) \mbox{\rm Im}[F_1^\ast G_1] +
(1-\hat{s}) \mbox{\rm Re}[F^\ast G] \Big] {\cal J}_{4} \Bigg\}~, \\ \nnb \\
\mbox{\rm where,} \nnb \\ \nnb \\
\label{e6722}
\Delta \es
16 m_B \hat{m}_\ell (1-\hat{s})^2
\Big( \mbox{\rm Im}[ (A_2^\ast + B_2^\ast) G_1] -
\mbox{\rm Re}[ (A_1^\ast + B_1^\ast) G - m_B \hat{m}_\ell 
(A_1^\ast B_1 + A_2^\ast B_2)] \Big) \nnb \\
\ar 48 m_B \hat{m}_\ell \hat{s} (1-\hat{s}) 
\mbox{\rm Im}[ (A_2^\ast + B_2^\ast) H_1] \nnb \\
\ek 8 m_B^3 \hat{m}_\ell \hat{s} (1-\hat{s})^2 
\mbox{\rm Im}[ (A_2^\ast + B_2^\ast) N_1] \nnb \\
\ar \frac{2}{3} (1-\hat{s})^2                                   
\Big[ 4 (3-v^2) \Big( \vel G_1 \ver^2 + \vel G \ver^2 \Big) +
m_B^2 \hat{s} (3+v^2) \Big( \vel A_1 \ver^2 + \vel A_2 \ver^2 +
\vel B_1 \ver^2 + \vel B_2 \ver^2 \Big) \Big] \nnb \\
\ar 16 \hat{s} v^2 \Big[ (1-\hat{s}) \mbox{\rm Re}[G^\ast H] + 
\hat{s} \vel H \ver^2 \Big] \nnb \\
\ar 16 \hat{s} (3-2 v^2) \Big[ (1-\hat{s}) \mbox{\rm Re}[G_1^\ast H_1] +            
\hat{s} \vel H_1 \ver^2 \Big] \nnb \\
\ek \frac{4}{3} m_B^2 \hat{s} (1-\hat{s})^2 (3-2 v^2) 
\Big( 2 \mbox{\rm Re}[G_1^\ast N_1] + m_B^2 \hat{s} 
\vel N_1 \ver^2 \Big) \nnb \\
\ek \frac{4}{3} m_B^2 \hat{s} (1-\hat{s})^2 v^2 
\Big( 2 \mbox{\rm Re}[G^\ast N] + m_B^2 \hat{s} \vel N \ver^2 \Big) \nnb \\
\ek \frac{1}{2} f_B^2 m_B^4 \vel F \ver^2 
\Big\{ (1-\hat{s})^2 v^2 ({\cal I}_1 + {\cal I}_{3}) -
(1+\hat{s}^2 + 2 \hat{s} v^2) {\cal I}_{2} -
[1-\hat{s} (4 - \hat{s} -2 v^2)] {\cal I}_{5} \Big\} \nnb \\
\ar \frac{1}{2} f_B^2 m_B^4 \vel F_1 \ver^2 \Big\{ - (1-\hat{s})^2 v^2 
({\cal I}_1 + {\cal I}_{3}) +
[1 - \hat{s} (2 - \hat{s} - 4 v^2 + 2 \hat{s} v^2 - 
2 \hat{s} v^4)] {\cal I}_{2} \nnb \\ 
\ek 2 \hat{s} (1-\hat{s}) v (1-v^2) {\cal I}_{4} +       
[1 - \hat{s} (2 - \hat{s} + 2 \hat{s} v^2 - 2 \hat{s} v^4)] {\cal I}_{5} \Big\} \nnb \\
\ek 4 f_B m_B^2 \hat{s} v \mbox{\rm Re} [F^\ast H] 
[(1-\hat{s}) v {\cal I}_{6} + (1+\hat{s}) {\cal I}_{7}] \nnb \\
\ek 4 f_B m_B^2 \hat{s} \mbox{\rm Im} [F_1^\ast H_1]
 [(1-\hat{s}) v^2 {\cal I}_{6} + 
(3 - 2 v^2 - 3 \hat{s} + 4 \hat{s} v^2) {\cal I}_{7}] \nnb \\
\ar 2 f_B m_B \hat{m}_\ell \mbox{\rm Re} [(A_1^\ast + B_1^\ast) F]
\Big[ 8 (1+\hat{s}) +
m_B^2 (1-\hat{s}^2) v^2 {\cal I}_{6} +
m_B^2 (1-\hat{s}) (1-3\hat{s}) {\cal I}_{7} \Big] \nnb \\
\ek f_B m_B \hat{m}_\ell (1-\hat{s}) 
\mbox{\rm Re}[(A_2^\ast + B_2^\ast) F_1] 
\Big[ 8 + m_B^2 (1 - 5 \hat{s}) v^2 {\cal I}_{6} +
m_B^2 (3 - 3\hat{s} + 4 \hat{s} v^2) {\cal I}_{7} \Big] \nnb \\
\ar f_B \mbox{\rm Im}[F_1^\ast G_1]
\Big[ -24 (1-\hat{s} + 2 \hat{s} v^2) +
m_B^2 (1-\hat{s}) (1+3 \hat{s} - 6 \hat{s} v^2) v^2 {\cal I}_{6} \nnb \\
\ek m_B^2 (1-\hat{s}) (1-\hat{s} - 2 \hat{s} v^2) {\cal I}_{7} \Big] \nnb \\
\ar f_B \mbox{\rm Re}[F^\ast G]  
\Big[ -24 (1+\hat{s}) +
m_B^2 (1-\hat{s}) (1-3 \hat{s}) v^2 {\cal I}_{6} -
m_B^2 (1-\hat{s}) (1-7 \hat{s} + 4 \hat{s} v^2) {\cal I}_{7} \Big] \nnb \\
\ar f_B m_B^2 \hat{s} \mbox{\rm Im}[F_1^\ast N_1]
\Big[ -8 (1-\hat{s} + 2 \hat{s} v^2) +
m_B^2 (1-\hat{s}) (3 + \hat{s} - 2 \hat{s} v^2) v^2 {\cal I}_{6} \nnb \\
\ar m_B^2 (1-\hat{s}) (3 - 2 v^2 - 3 \hat{s} + 4 \hat{s} v^2 ) 
{\cal I}_{7} \Big] \nnb \\
\ar f_B m_B^2 \hat{s} \mbox{\rm Re}[F^\ast N]  
\Big[ -8 (1+\hat{s}) +
m_B^2 (1-\hat{s}) (3 - \hat{s}) v^2 {\cal I}_{6} +
m_B^2 (1-\hat{s}^2) {\cal I}_{7} \Big]~.
\eea

In Eqs. (\ref{e6716})--(\ref{e6722}), $\hat{s} = q^2/m_B^2$, 
$v=\sqrt{1-4 \hat{m}_\ell^2/\hat{s}}$ is the lepton
velocity with $\hat{m}_\ell = m_\ell/m_B$, and 
${\cal I}_i$ represent the following integrals
\bea
{\cal I}_i \es \int_{-1}^{+1} {\cal F}_i(z) dz~,\nnb \\
{\cal J}_i \es \int_{0}^{+1} {\cal G}_i(z) dz - 
               \int_{-1}^{0} {\cal G}_i(z) dz ~,\nnb
\eea
where
\bea
\begin{array}{lll}
{\cal G}_{1} = \ds\frac{z \sqrt{1-z^2}}{(p_1 \cdot k) (p_2 \cdot k)}~,&
{\cal G}_{2} = \ds\frac{z \sqrt{1-z^2}}{(p_1 \cdot k)^2}~,&
{\cal G}_{3} = \ds\frac{\sqrt{1-z^2}}{(p_1 \cdot k)^2}~,\\ \\
{\cal G}_{4} = \ds\frac{z \sqrt{1-z^2}}{(p_1 \cdot k)}~,&    
{\cal F}_{1} = \ds\frac{z^2}{(p_1 \cdot k) (p_2 \cdot k)}~,&    
{\cal F}_{2} = \ds\frac{1}{(p_1 \cdot k) (p_2 \cdot k)}~,\\ \\
{\cal F}_{3} = \ds\frac{z^2}{(p_1 \cdot k)^2}~,&
{\cal F}_{4} = \ds\frac{z}{(p_1 \cdot k)^2}~,&
{\cal F}_{5} = \ds\frac{1}{(p_1 \cdot k)^2}~,\\ \\
{\cal F}_{6} = \ds\frac{z^2}{p_1 \cdot k}~,&
{\cal F}_{7} = \ds\frac{1}{p_1 \cdot k}~.
\end{array} \nnb
\eea

We note that, the forward--backward asymmetries ${\cal A}_{NN}$, 
${\cal A}_{NT}$, ${\cal A}_{TN}$ and ${\cal A}_{TT}$ are all equal to zero.

\section{Numerical analysis and discussion}

In this section we present our numerical analysis for all possible
polarized forward--backward asymmetries of leptons. The values of the 
input parameters which we have used in the numerical analysis are:
$\vel V_{tb} V_{ts}^\ast \ver = 0.0385$, $m_\mu=0.106~GeV$, $m_\tau=1.78~GeV$, 
$m_b=4.8~GeV$. For the SM values of the Wilson coefficients
we have used $C_7^{SM}(m_b)=-0.313$, $C_9^{SM}(m_b) = 4.344$ and
$C_{10}^{SM}(m_b) = -4.669$. The magnitude of $C_7^{SM}$ is quite well
determined from the $b \rar s \gamma$ transition, and hence it is well 
established. Therefore the values of $C_{BR}$ and $C_{SL}$ are fixed by the 
relations $C_{BR}=-2 m_b C_7^{eff}$ and $C_{SL}=-2 m_s C_7^{eff}$. It is well 
known that the Wilson coefficient $C_9^{SM}$ receives also long distance
contributions which have their origin in the real $\bar{c}c$ intermediate
states, i.e., $J/\psi$, $\psi^\prime$, $\cdots$ \cite{R6715}. In the present
work we consider only short distance contributions.

The values of the new Wilson coefficients are needed in order to carry out
the numerical calculations for ${\cal A}_{ij}$ given in Eqs. (\ref{e6717})--
(\ref{e6722}). All new Wilson coefficients are varied in the range 
$- \vel C_{10}^{SM} \ver \le C_X \le \vel C_{10}^{SM} \ver$ and it is
assumed that they are real. The experimental results on the branching ratio of the
$B_s \rar K^\ast (K) \ell^+ \ell^-$ decays  \cite{R6716,R6717} and the bound 
on the branching ratio of $B_s \rar \mu^+ \mu^-$ \cite{R6718} suggest that
this is the right order of magnitude for the Wilson coefficients describing
the vector and scalar interaction coefficients. But present experimental
results on the branching ratio of the 
$B_s \rar K^\ast \ell^+ \ell^-$ and $B_s \rar K \ell^+ \ell^-$ decays impose
stronger restrictions on some of the new Wilson coefficients. For example,
$-2 \le C_{LL} \le 0$, $0 \le C_{RL} \le 2.3$, $-1.5 \le C_{T} \le 1.5$ and
$-3.3 \le C_{TE} \le 2.6$, and all of the remaining Wilson coefficients vary
in the region $- \vel C_{10}^{SM} \ver \le C_X \le \vel C_{10}^{SM} \ver$.

It follows from the expressions of all forward--backward asymmetries of the
leptons that, explicit forms of the form factors are needed,
which are the main and most important parameters in the calculation of
${\cal A}_{ij}$. These form factors are calculated in the framework of the QCD sum
rules in \cite{R6703,R6713,R6714} whose $q^2$ dependences are given as
\bea
\begin{array}{ll}
g_(q^2) = \ds\frac{1~GeV}{\ds \ga 1-\frac{q^2}{(5.6~GeV)^2}\dr^2}~,&
f_(q^2) = \ds\frac{0.8~GeV}{\ds \ga 1-\frac{q^2}{(6.5~GeV)^2}\dr^2}~,\\ \\  
g_1(q^2) = \ds\frac{3.74~GeV^2}{\ds \ga 1-\frac{q^2}{40.5~GeV^2}\dr^2}~,&
f_1(q^2) = \ds\frac{0.67~GeV^2}{\ds \ga 1-\frac{q^2}{30~GeV^2}\dr^2}~,
\end{array} \nnb
\eea
which we will use in the numerical analysis.

Numerical results are presented only for the $B_s \rar \ell^+ \ell^- \gamma$  
decay, because in the SU(3) limit the difference between the decay rates of
$B_s \rar \ell^+ \ell^- \gamma$  and $B_d \rar \ell^+ \ell^- \gamma$ is
attributed only to the CKM matrix elements. In other words, the decay rate
of the $B_s \rar \ell^+ \ell^- \gamma$ is approximately 20 times larger
compared to that of decay rate of $B_d \rar \ell^+ \ell^- \gamma$, that is
\bea
\frac{\Gamma(B_d \rar \ell^+ \ell^- \gamma)}
{\Gamma(B_s \rar \ell^+ \ell^- \gamma)} \simeq \frac{\vel V_{tb}
V_{td}^\ast\ver^2}
{\vel V_{tb} V_{ts}^\ast\ver^2} \simeq \frac{1}{20}~.\nnb
\eea 

We now proceed by commenting on the result of our numerical analysis.
Firstly, we study the dependence of the polarized forward--backward
asymmetries on $q^2$ at five different values of the new Wilson
coefficients. Our detailed numerical analysis shows that for the 
$B_s \rar \mu^+ \mu^- \gamma$  decay only the ${\cal A}_{FB}^{LT}$ and 
${\cal A}_{FB}^{TL}$ asymmetries have zero positions (the numerical values
of the asymmetries ${\cal A}_{FB}^{LN}$ and ${\cal A}_{FB}^{NL}$ are very
small and hence we do not present them). In Fig. (1) we present the
dependence of ${\cal A}_{FB}^{LT}$ on $q^2$ at five fixed values of the
scalar interaction coefficient $C_{LRLR} = -4;-2;0;+2;+4$. From this figure
we see that the zero position which occurs for positive values of
$C_{LRLR}$ is shifted to right for increasing values of $C_{LRLR}$. The same
figure also depicts that the zero position of ${\cal A}_{FB}^{LT}$ is absent
for the SM case. Therefore, determination of the zero position of ${\cal
A}_{FB}^{LT}$ is an unambiguous indication of the new physics beyond the SM,
as well as allowing us determine the sign of the scalar interaction
coefficients $C_{LRLR}$. In Fig. (2) we present the dependence of ${\cal
A}_{FB}^{LT}$ on $q^2$ at fixed values of $C_{RLLR}$. Similar to the
previous case, zero position of the ${\cal A}_{FB}^{LT}$ appears again, but
the difference from it being it occurs for the negative values of
$C_{RLLR}$. It should be noted here that the zero position of ${\cal
A}_{FB}^{LT}$ is present for the remaining scalar interaction coefficients 
$C_{LRRL}$ and $C_{RLRL}$ as well, which can be seen in Figs. (3) and (4).
More interesting observation for these cases is that, the zero position
appears for $q^2 < 2~GeV^2$ and hence it is free of the long distance
$J/\psi$ contributions. As far as $B_s \rar \mu^+ \mu^- \gamma$   
decay is concerned, our numerical analysis shows that the zero position of 
${\cal A}_{FB}^{LT}$ is absent for all Wilson coefficients other than the 
scalar interaction coefficients. Hence, determination of the zero position 
of ${\cal A}_{FB}^{LT}$ can serve as a good test for establishing new 
physics beyond the SM due to the presence of the scalar interaction 
coefficients.

The situation for the ${\cal A}_{FB}^{TL}$ asymmetry for the  $B_s \rar \mu^+
\mu^- \gamma$  decay is richer in content compared to that of the ${\cal
A}_{FB}^{LT}$ case. For this forward--backward asymmetry, the zero position
occurs for all new Wilson coefficients. In Figs. (5)--(11) we present the
dependence of ${\cal A}_{FB}^{TL}$ on $q^2$ at five fixed values of the new
Wilson coefficients. These figures depict that:

\begin{itemize}

\item For vector interactions with the Wilson coefficients $C_{LL}$ and
$C_{RR}$, the zero position of ${\cal A}_{FB}^{TL}$ is shifted. When these
coefficients get positive (negative) values, the zero position of ${\cal
A}_{FB}^{TL }$ is shifted to the left (right) compared to that of the SM
case. In the presence of the Wilson coefficients $C_{LR}$ and $C_{RL}$ the
zero position of the ${\cal A}_{FB}^{TL}$ is shifted to the right (left)
compared to that of the SM result, when these Wilson coefficients are
positive (negative).

\item In the presence of the scalar interactions with the coefficients 
$C_{LRRL}$ and $C_{RLRL}$, the zero position of ${\cal A}_{FB}^{TL}$ is
shifted the left compared to that of the SM result. The zero position for
$C_{LRRL}$ occurs only for its positive values, while it occurs only for the
negative values of $C_{RLRL}$. 

In the presence of scalar interactions
$C_{RLLR}$ and $C_{LRLR}$, no new zero position of ${\cal A}_{FB}^{TL}$ occurs
with respect to the one for the SM case.

\item New zero positions of ${\cal A}_{FB}^{TL}$ are observed in the
presence of the tensor interaction for the positive values of $C_{T}$, and
the zero position is shifted to the left,    

\end{itemize}

In the case of $B_s \rar \tau^+ \tau^- \gamma$    
decay, similar to the $B_s \rar \mu^+ \mu^- \gamma$   
decay, we observe that several of the polarized
forward--backward asymmetries are very sensitive to the existence of new
physics. Let us briefly summarize our results:

i) Among all polarization asymmetries (which can be measurable in the
experiments) only ${\cal A}_{FB}^{TL}$ is very sensitive to the existence of
all types of new physics interactions, except to the presence of the vector
interactions with coefficients $C_{LL}$ and $C_{RR}$.

ii) ${\cal A}_{FB}^{LL}$ is sensitive to the presence of the tensor
interaction and its zero position occurs for $C_{T}=+4$ at $q^2\approx
17~GeV^2$, while zero position of ${\cal A}_{FB}^{LL}$ is absent for the SM case.
Therefore, determination of the zero position of ${\cal A}_{FB}^{LL}$ can
confirm the existence of the tensor interaction in the $B_s \rar \tau^+
\tau^- \gamma$  decay (see Fig. (12)).

\begin{itemize}

\item ${\cal A}_{FB}^{TL}$ exhibits similar dependence on $C_{LR}$ and
$C_{RL}$. The zero position of ${\cal A}_{FB}^{TL}$ is shifted to to the
left (right) when $C_{LR}$ and $C_{RL}$ are negative (positive) compared to
that of the SM prediction. Note that the zero position of ${\cal
A}_{FB}^{TL}$ lies on the left side for the vector interaction $C_{LR}$
compared to the zero position of the $C_{RL}$  (see Figs. (13), (14)).

\item ${\cal A}_{FB}^{TL}$ shows stronger dependence on the scalar
interactions $C_{LRRL}$ and $C_{RLRL}$. The magnitude of ${\cal
A}_{FB}^{TL}$ increases (decreases) as the new Wilson coefficient $C_{LRRL}$
gets positive (negative) values. This behavior is to the contrary for the
coefficient $C_{RLRL}$ (see Figs. (15), (16)).

\item In the presence of the tensor interaction with the coefficient
$C_{T}$, zero position of the asymmetry ${\cal A}_{FB}^{TL}$ is located on
the left side of the SM prediction for negative values of $C_{T}$ (Fig. (17)). 

\end{itemize}
  
We see from the explicit expressions of the polarized forward--backward
asymmetries that they all depend both on $q^2$ and the new Wilson
coefficients. For this reason there may appear difficulties in the
experiments in studying the dependence of the physical observables on both
parameters simultaneously. In order to get "pure information" about about
new physics, we eliminate the dependence of physical quantities on $q^@$, by
performing integration over $q^2$ in the kinematically allowed region, i.e.,
we average the polarized forward backward asymmetry
\bea
\label{e6726}
\lla {\cal A}_{ij} \rra = \ds \frac{\int_{4 m_\ell^2}^{m_B^2}
{\cal A}_{ij} \ds \frac{d {\cal B}}{dq^2} dq^2}
{\int_{4 m_\ell^2}^{m_B^2}
\ds \frac{d {\cal B}}{dq^2} dq^2}~. \nnb
\eea             

In Fig. (18) we depict the dependence of $\lla {\cal A}_{FB}^{LL} \rra$ 
on the new Wilson for the $B_s \rar \mu^+ \mu^- \gamma$  decay. From this 
figure we see that $\lla {\cal A}_{FB}^{LL} \rra$ shows symmetric behavior 
in its dependence on all scalar interactions; and except for regions
$-4 < C_{RR},C_{RL} < 0$, $-0.4 < C_{LRRL},C_{LRRL}< 0$ and $0 \le
C_{RLRL},C_{LRLR}< 0.4$ it is larger compared to the SM result (SM result
corresponds to the intersection point of all curves). It is also interesting
to observe that  $\lla {\cal A}_{FB}^{LL} \rra >  \lla {\cal A}_{FB}^{SM}
\rra$ for only negative values of $C_{RR}$. 

Our numerical analysis furthers shows that, for the $B_s \rar \mu^+ \mu^-
\gamma$  decay, $\lla {\cal A}_{FB}^{LT} \rra$ is sensitive only to $C_{T}$
and at negative (positive) values of $C_{T}$ $\lla {\cal A}_{FB}^{LT} \rra$           
is positive (negative) and larger (smaller) compared to the SM result.
Therefore, determination of the sign and magnitude of $\lla {\cal
A}_{FB}^{LT} \rra$ can serve as a good test for establishing existence
of the tensor interaction.

The dependence of $\lla {\cal A}_{FB}^{TL} \rra$ on the new Wilson
coefficients for the $B_s \rar \mu^+ \mu^- \gamma$  decay is presented in Fig.
(19). We observe from this figure that $\lla {\cal A}_{FB}^{TL} \rra$
shows stronger dependence on the tensor interaction coefficient $C_{T}$ and
scalar interactions $C_{RLRLR}$ and $C_{LRRL}$. 

In Figs. (20), (21), and (22) we present the dependence of $\lla {\cal
A}_{FB}^{LL} \rra$, $\lla {\cal A}_{FB}^{LT} \rra$ and $\lla {\cal A}_{FB}^{TL}
\rra$ on the new Wilson coefficients for the $B_s \rar \tau^+ \tau^- \gamma$  
decay, respectively. Fig. (20) depicts that $\lla {\cal A}_{FB}^{LL} \rra$
exhibits considerable departure from the SM result for the scalar
interactions and the vector interaction with coefficient $C_{RR}$. We see
from Fig. (21) that when new Wilson coefficients are negative $\lla {\cal
A}_{FB}^{LT} \rra$ shows stronger dependence on on the tensor interaction 
($C_{T}$) and scalar type interactions, and when $C_X > 0$, 
$\lla {\cal A}_{FB}^{LT} \rra$ exhibits strong dependence on vector
interactions and the tensor interaction with the coefficient $C_{TE}$.

At the end of this section, we discuss the problem of the detectability of
forward--backward asymmetry in the experiments. Experimentally, to measure
an asymmetry $\la {\cal A}_{ij} \ra$ of the decay with the branching ratio 
${\cal B}$ at $n \sigma$ level, the required number of events
(i.e., the number of $B \bar{B}$ pair) are given by
\bea
{\cal N} = \frac{n^2}{{\cal B} s_1 s_2 \la {\cal A}_{ij} \ra^2}~,\nnb
\eea
where $s_1$ and $s_2$ are the efficiencies of the leptons. Efficiency of 
the $\mu$ lepton is practically equal to one, and typical values of the
efficiency of the $\tau$ lepton ranges from $50\%$ to $90\%$ for
the various decay modes \cite{R6719}.

From the expression for ${\cal N}$ we see that, in order to obtain the 
forward--backward asymmetries in $B_s \rar \ell^+ \ell^- \gamma$  
decays at $3\sigma$ level, the minimum number of required events
are (for the efficiency of $\tau$--lepton we take $0.5$, and for 
$\la {\cal A}_{ij} \ra$, their maximal values beyond the SM):

\begin{itemize}
\item for the $B_s \rar \mu^+ \mu^- \gamma$  decay
\bea
{\cal N} = \left\{ \begin{array}{ll}
\sim 2 \times 10^{9}     & \lla {\cal A}_{LL}\rra,\\
\sim 3 \times 10^{10}  & \lla {\cal A}_{LT} \rra \simeq \lla {\cal A}_{TL}\rra,
\end{array} \right. \nnb
\eea
which yields that, for detecting $\lla {\cal
A}_{LT} \rra$ and $\lla {\cal A}_{TL} \rra$, more than $10^{13}$ $\bar{B} B$
pairs are required.

\item for $B_s \rar \tau^+ \tau^- \gamma$  decay
\bea
{\cal N} = \sim 6 \times 10^{11}   \lla {\cal A}_{LL}\rra,~
\lla {\cal A}_{LT} \rra,~\lla {\cal A}_{TL}\rra. \nnb
\eea
\end{itemize}

The number of $\bar{B} B$ pairs that will be produced at LHC is
around $\sim 10^{12}$. As a result of a comparison of this number of
$\bar{B} B$ pairs with that of ${\cal N}$, we conclude that $\lla {\cal
A}_{LL} \rra$, $\lla {\cal A}_{TL} \rra$ and $\lla {\cal A}_{TL} \rra$
in both decays can be detectable in "beyond the SM
scenarios" in future experiments at LHC. Note that
in the SM, only $\lla {\cal A}_{LL} \rra$ for the $B_s \rar \mu^+ \mu^-
\gamma$  decay can be detectable at LHC. Therefore, observation of these
asymmetries can be explained only by new physics beyond the SM.

In conclusion, we calculate polarized forward--backward asymmetries
using the most general, model independent form of the effective Hamiltonian
including all possible form of interactions. The sensitivity of the
averaged polarized forward--backward asymmetries to the new Wilson
coefficients are studied. Finally we discuss the possibility of experimental
measurement of these double--lepton polarization asymmetries at LHC.

\newpage

\section*{Figure captions}

{\bf Fig. (1)} The dependence of the polarized forward--backward asymmetry 
${\cal A}_{FB}^{LT}$ on $q^2$ at four fixed values of $C_{LRLR}$ for the
$B_s \rar \mu^+ \mu^- \gamma$  decay.\\ \\
{\bf Fig. (2)} The same as in Fig. (1), but for at four fixed values of 
$C_{RLLR}$.\\ \\
{\bf Fig. (3)} The same as in Fig. (1), but for at four fixed values of 
$C_{LRRL}$.\\ \\
{\bf Fig. (4)} The same as in Fig. (1), but for at four fixed values of
$C_{RLRL}$.\\ \\ 
{\bf Fig. (5)} The dependence of the polarized forward--backward asymmetry
${\cal A}_{FB}^{TL}$ on $q^2$ at four fixed values of $C_{LL}$ for the
$B_s \rar \mu^+ \mu^- \gamma$  decay.\\ \\
{\bf Fig. (6)} The same as in Fig. (5), but for at four fixed values of
$C_{LR}$.\\ \\
{\bf Fig. (7)} The same as in Fig. (5), but for at four fixed values of
$C_{RL}$.\\ \\
{\bf Fig. (8)} The same as in Fig. (5), but for at four fixed values of
$C_{RR}$.\\ \\
{\bf Fig. (9)} The same as in Fig. (5), but for at four fixed values of
$C_{LRRL}$.\\ \\
{\bf Fig. (10)} The same as in Fig. (5), but for at four fixed values of
$C_{RLRL}$.\\ \\
{\bf Fig. (11)} The same as in Fig. (5), but for at four fixed values of
$C_{T}$.\\ \\
{\bf Fig. (12)} The dependence of the polarized forward--backward asymmetry
${\cal A}_{FB}^{LL}$ on $q^2$ at four fixed values of $C_{T}$ for the
$B_s \rar \tau^+ \tau^- \gamma$  decay.\\ \\
{\bf Fig. (13)} The dependence of the polarized forward--backward asymmetry
${\cal A}_{FB}^{TL}$ on $q^2$ at four fixed values of $C_{LR}$ for the
$B_s \rar \tau^+ \tau^- \gamma$  decay.\\ \\
{\bf Fig. (14)} The same as in Fig. (13), but for at four fixed values of
$C_{RL}$.\\ \\
{\bf Fig. (15)} The same as in Fig. (13), but for at four fixed values of
$C_{LRRL}$.\\ \\
{\bf Fig. (16)} The same as in Fig. (13), but for at four fixed values of
$C_{RLRL}$.\\ \\
{\bf Fig. (17)} The same as in Fig. (13), but for at four fixed values of
$C_{T}$.\\ \\
{\bf Fig. (18)} The dependence of the polarized forward--backward asymmetry
${\cal A}_{FB}^{LL}$ on the new Wilson coefficients for the 
$B_s \rar \mu^+ \mu^- \gamma$  decay.\\ \\
{\bf Fig. (19)} The same as in Fig. (18), but for the polarized
forward--backward asymmetry ${\cal A}_{FB}^{TL}$.\\ \\
{\bf Fig. (20)} The same as in Fig. (18), but for the 
$B_s \rar \tau^+ \tau^- \gamma$  decay.\\ \\
{\bf Fig. (21)} The same as in Fig. (20), but for the polarized
forward--backward asymmetry ${\cal A}_{FB}^{LT}$.\\ \\
{\bf Fig. (22)} The same as in Fig. (21), but for the polarized
forward--backward asymmetry ${\cal A}_{FB}^{TL}$.\\ \\

\newpage

\begin{figure}
\vskip 1.5 cm
    \includegraphics{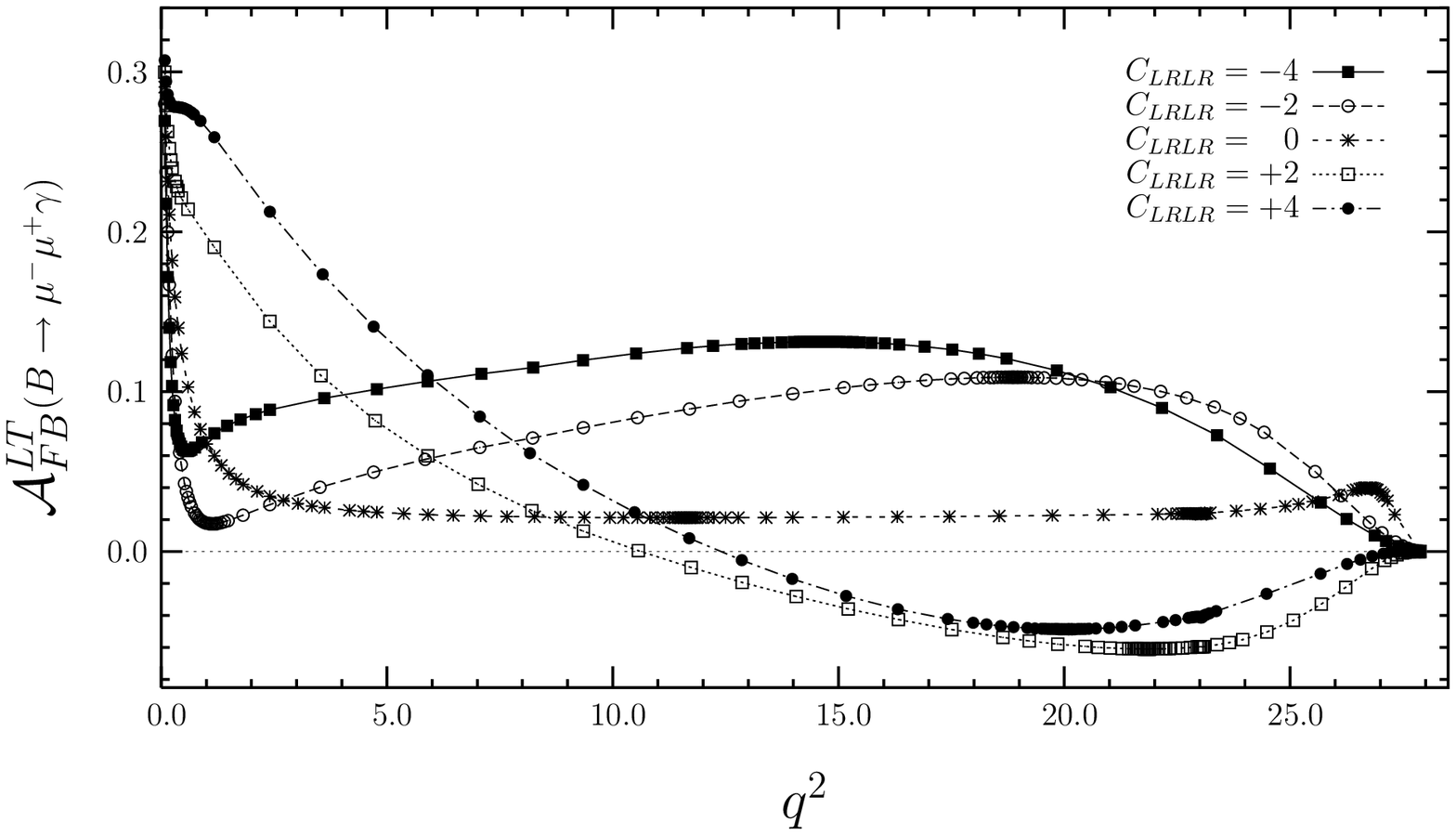}
\vskip 7.8cm
\caption{}
\end{figure}

\begin{figure}
\vskip 2.5 cm
    \includegraphics{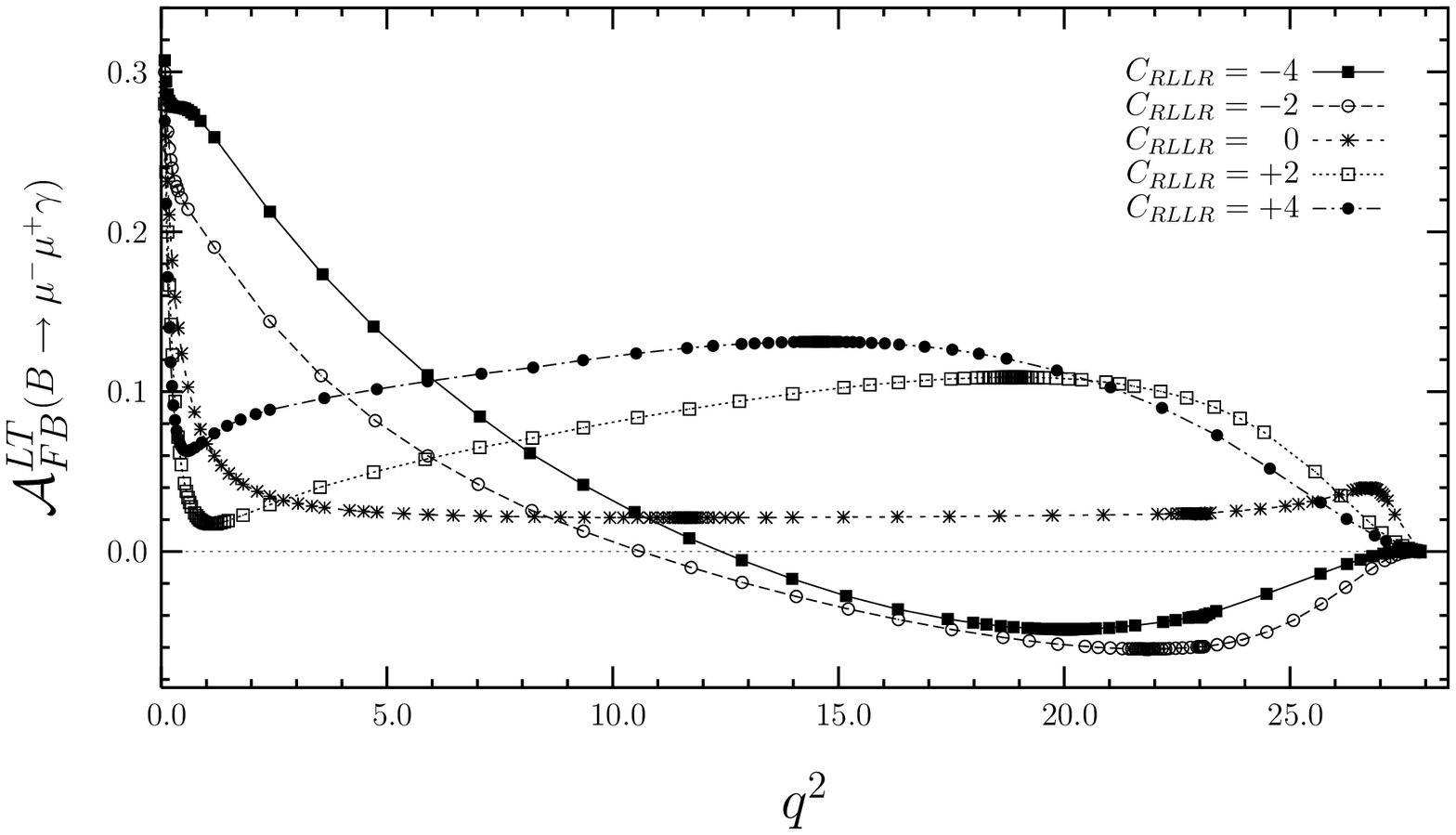}
\vskip 7.8 cm
\caption{}
\end{figure}

\begin{figure}
\vskip 1.5 cm
    \includegraphics{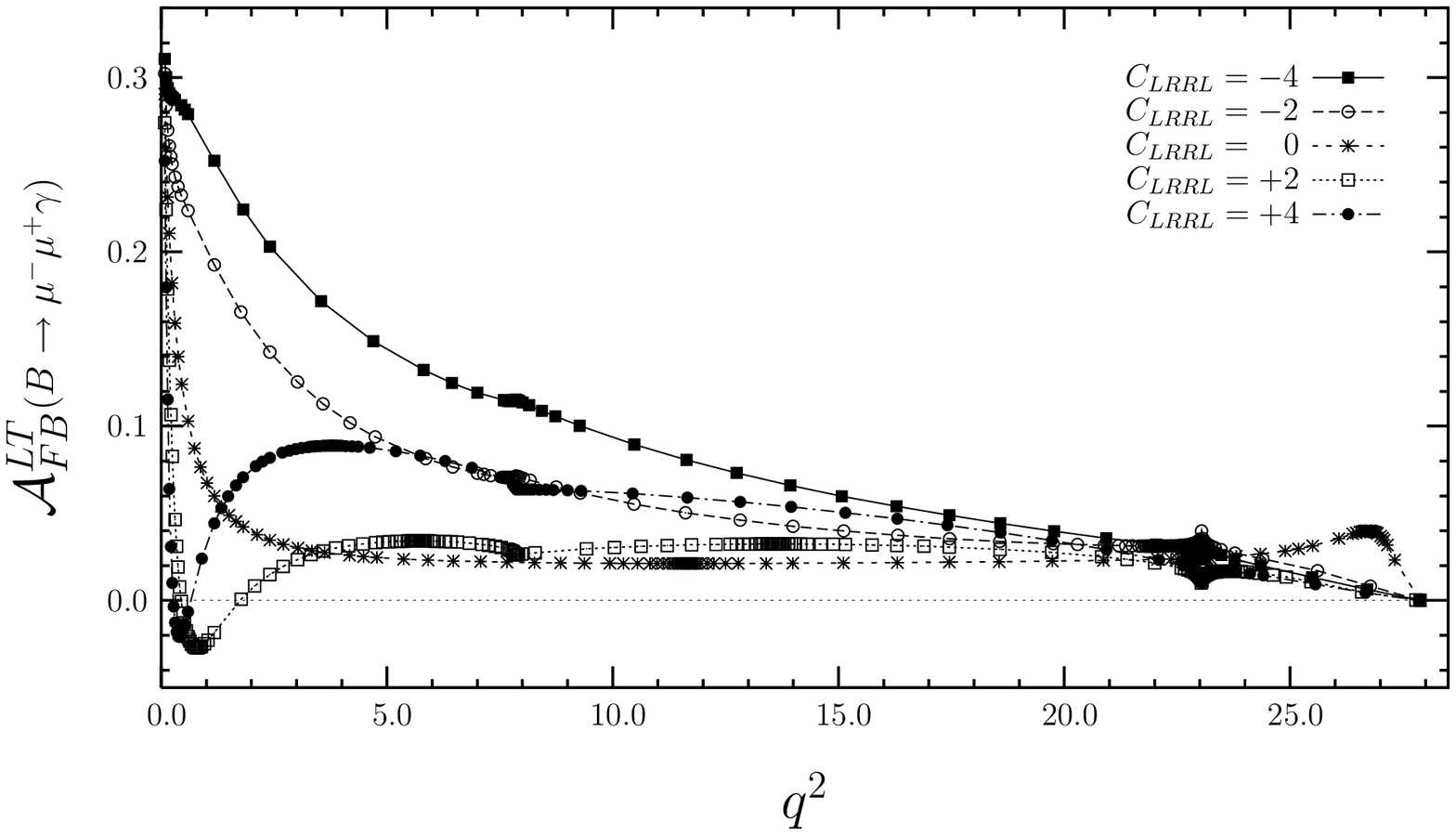}
\vskip 7.8cm
\caption{}
\end{figure}

\begin{figure}
\vskip 2.5 cm
    \includegraphics{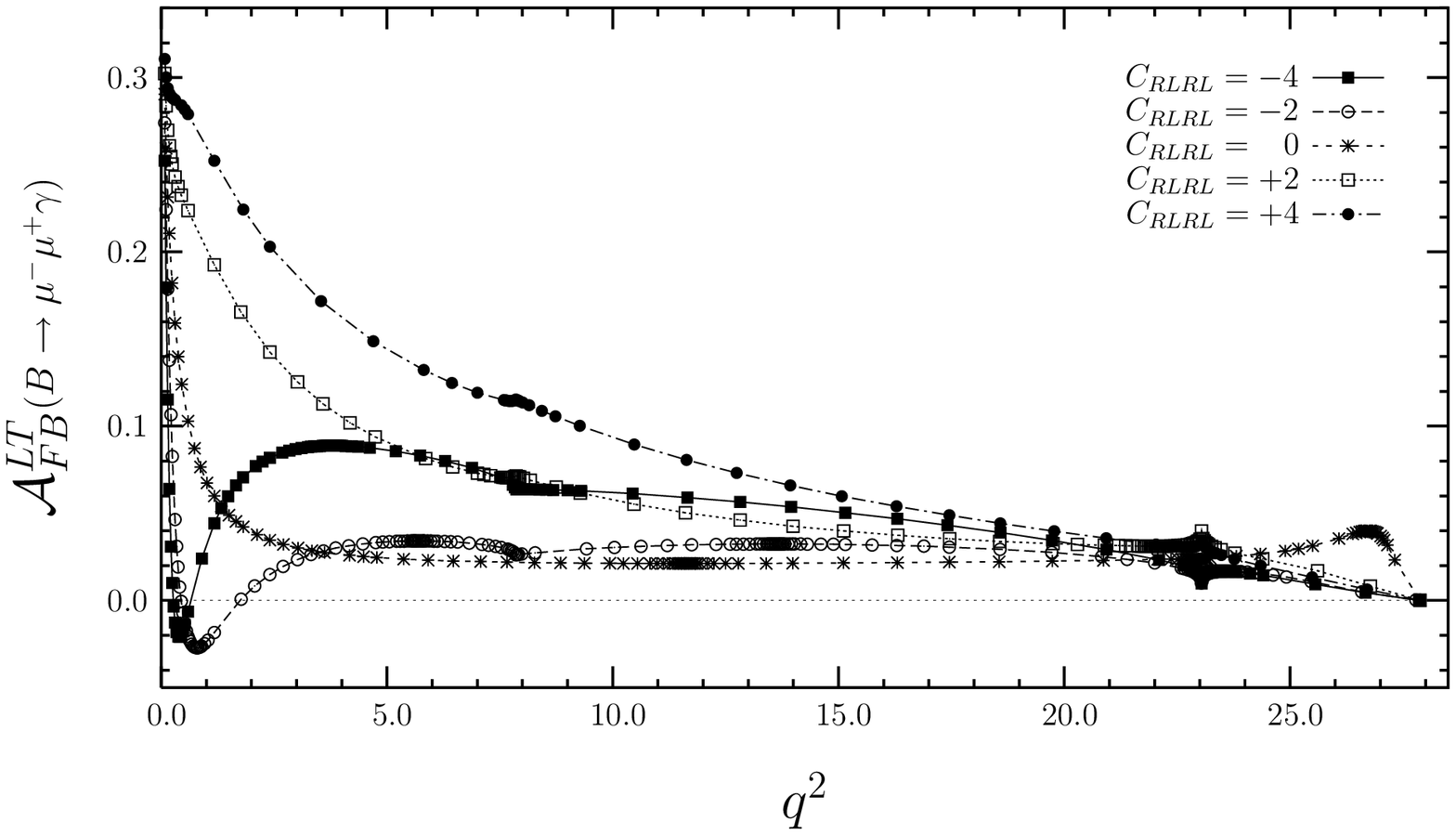}
\vskip 7.8 cm
\caption{}
\end{figure}

\begin{figure}
\vskip 2.5 cm
    \includegraphics{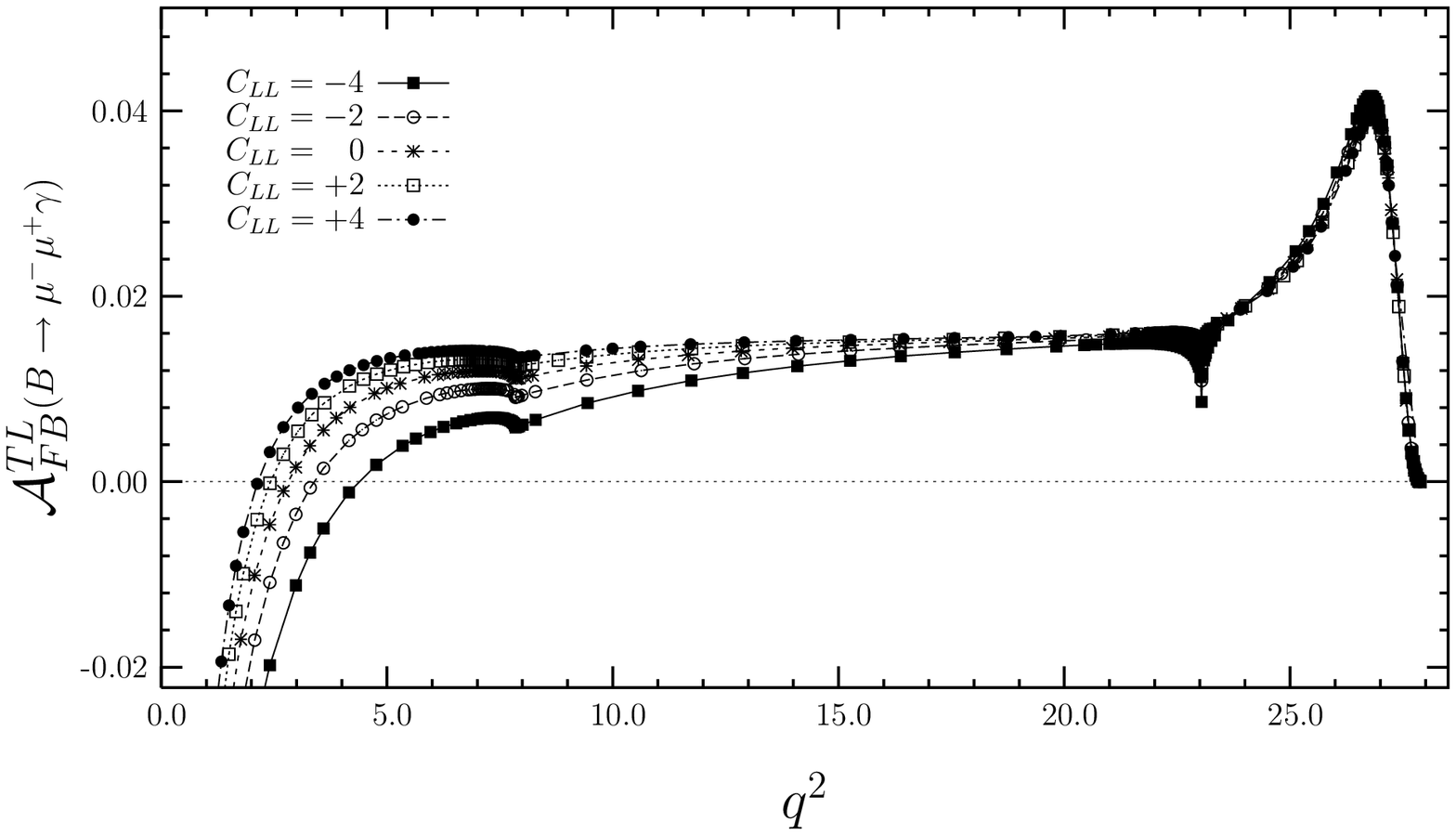}
\vskip 7.8 cm
\caption{}
\end{figure}

\begin{figure}
\vskip 1.5 cm
    \includegraphics{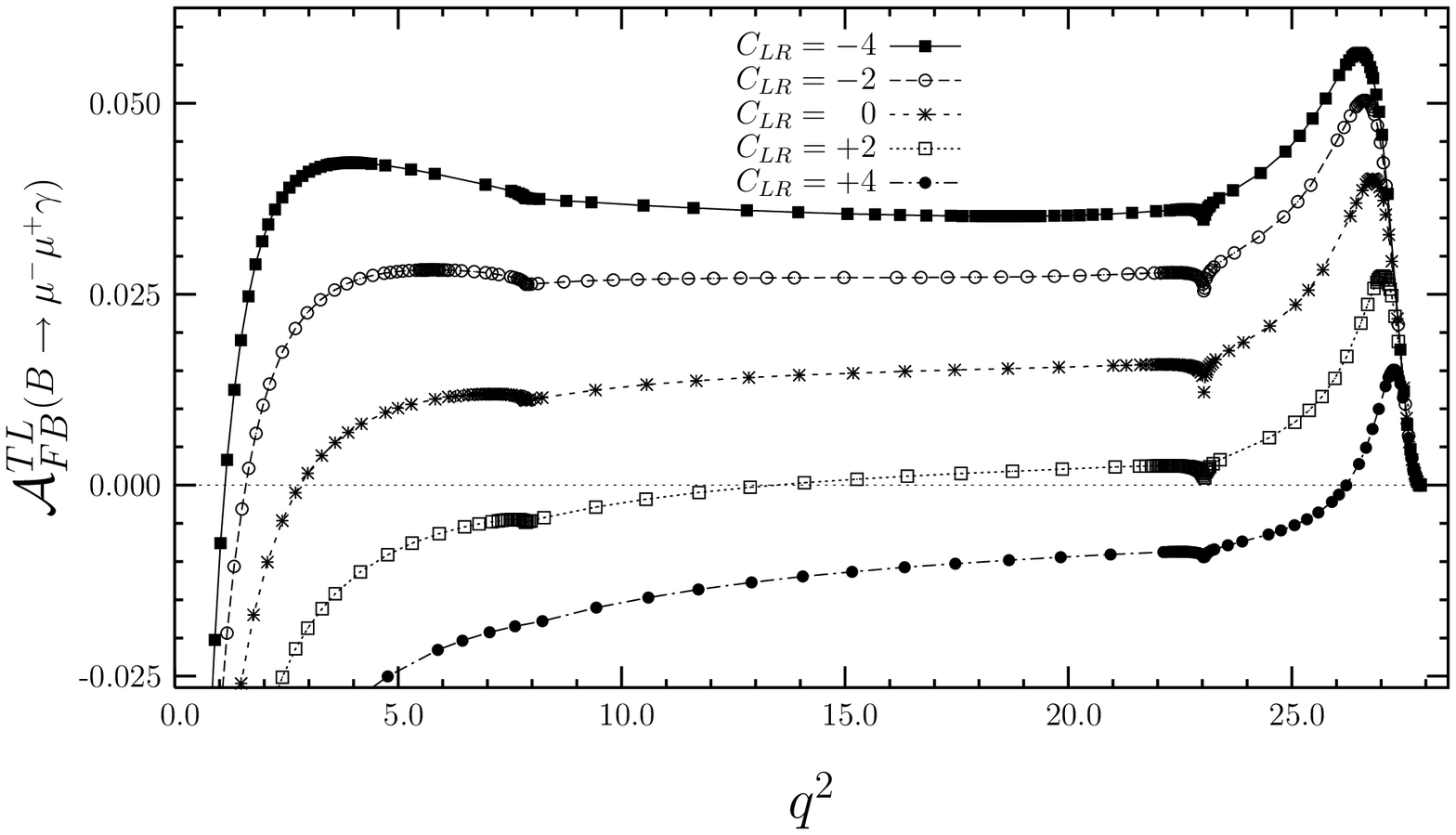}
\vskip 7.8cm
\caption{}
\end{figure}

\begin{figure}
\vskip 2.5 cm
    \includegraphics{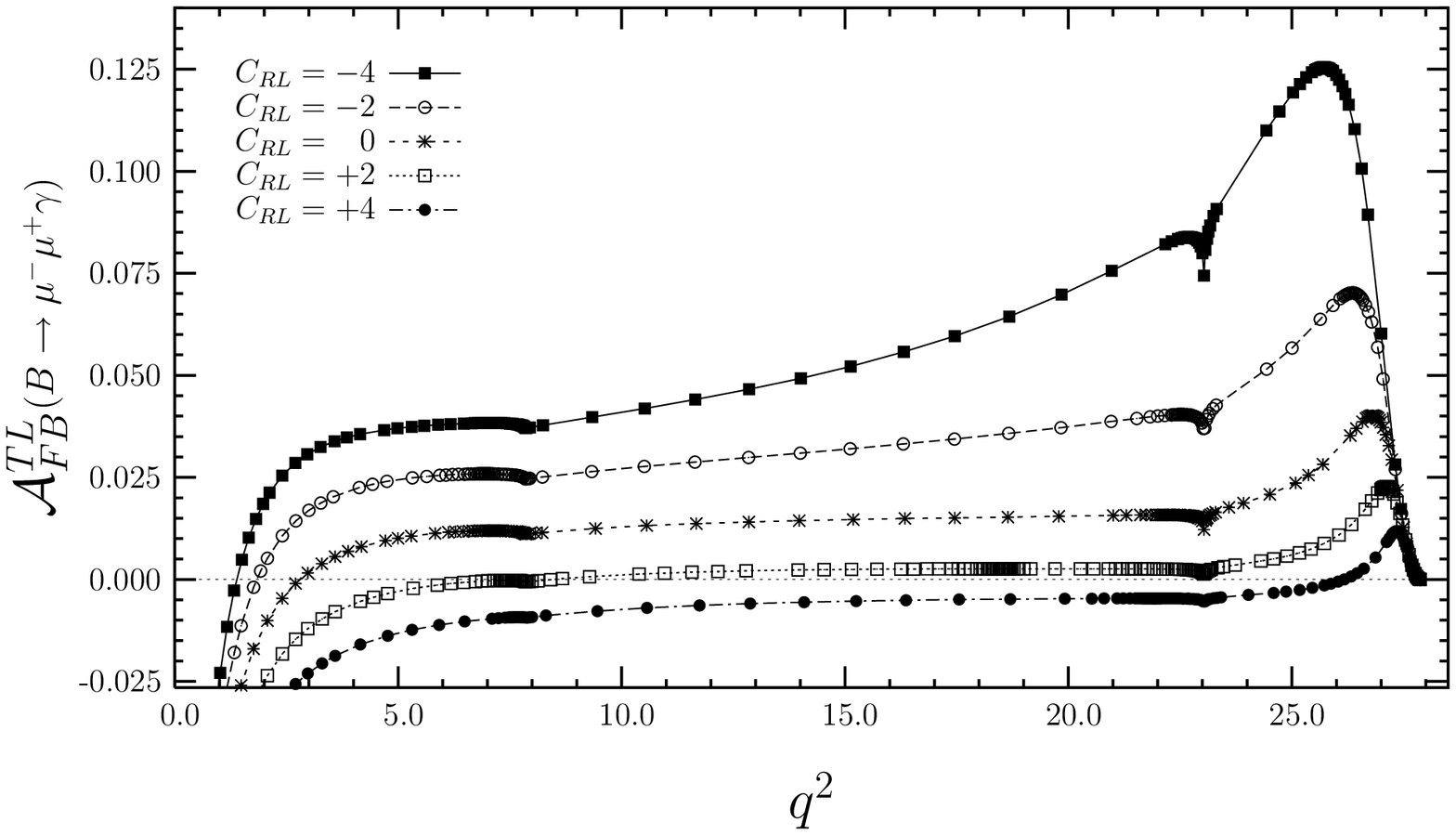}
\vskip 7.8 cm
\caption{}
\end{figure}

\begin{figure}
\vskip 2.5 cm
    \includegraphics{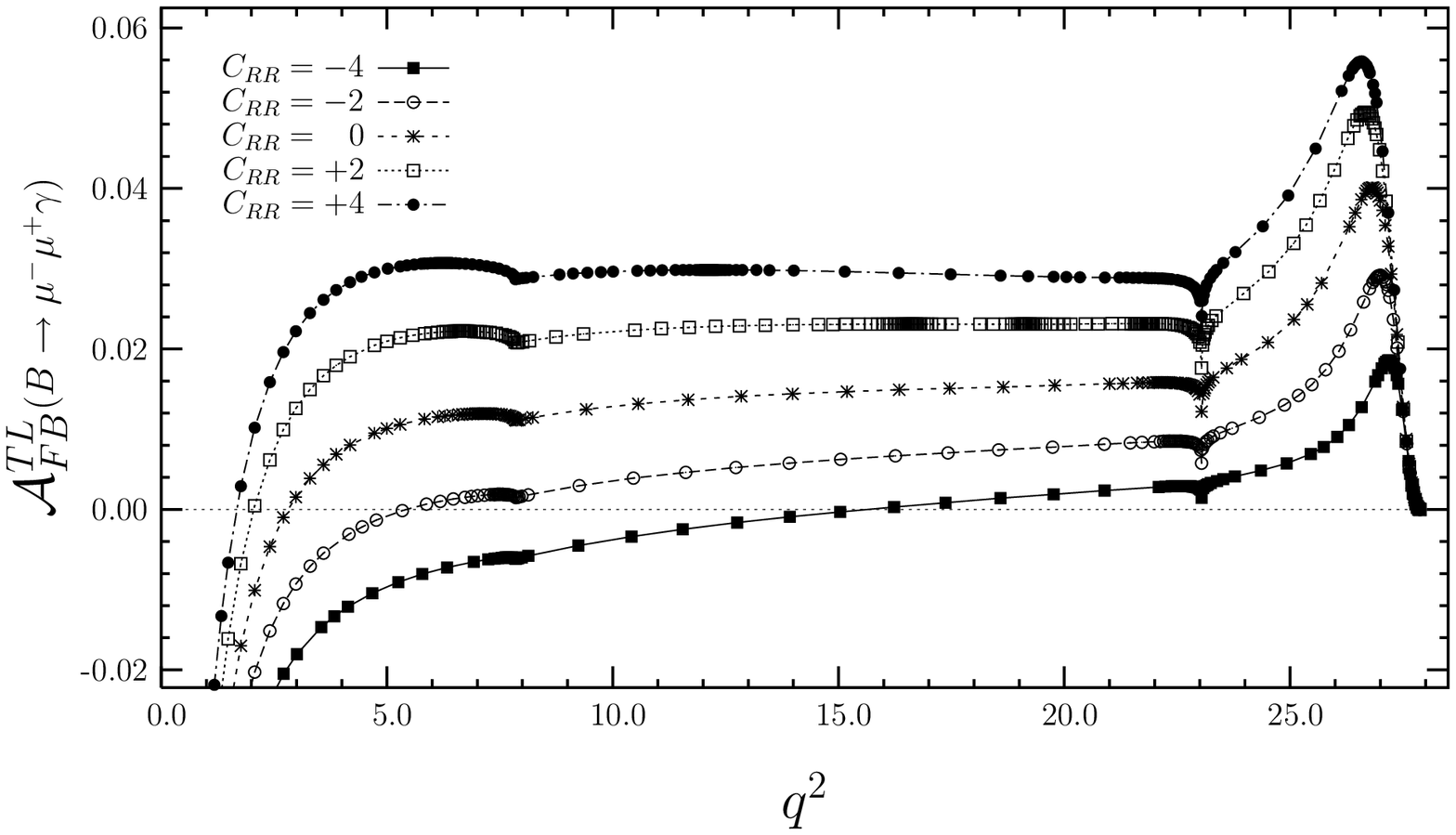}
\vskip 7.8 cm
\caption{}
\end{figure}

\begin{figure}
\vskip 2.5 cm
    \includegraphics{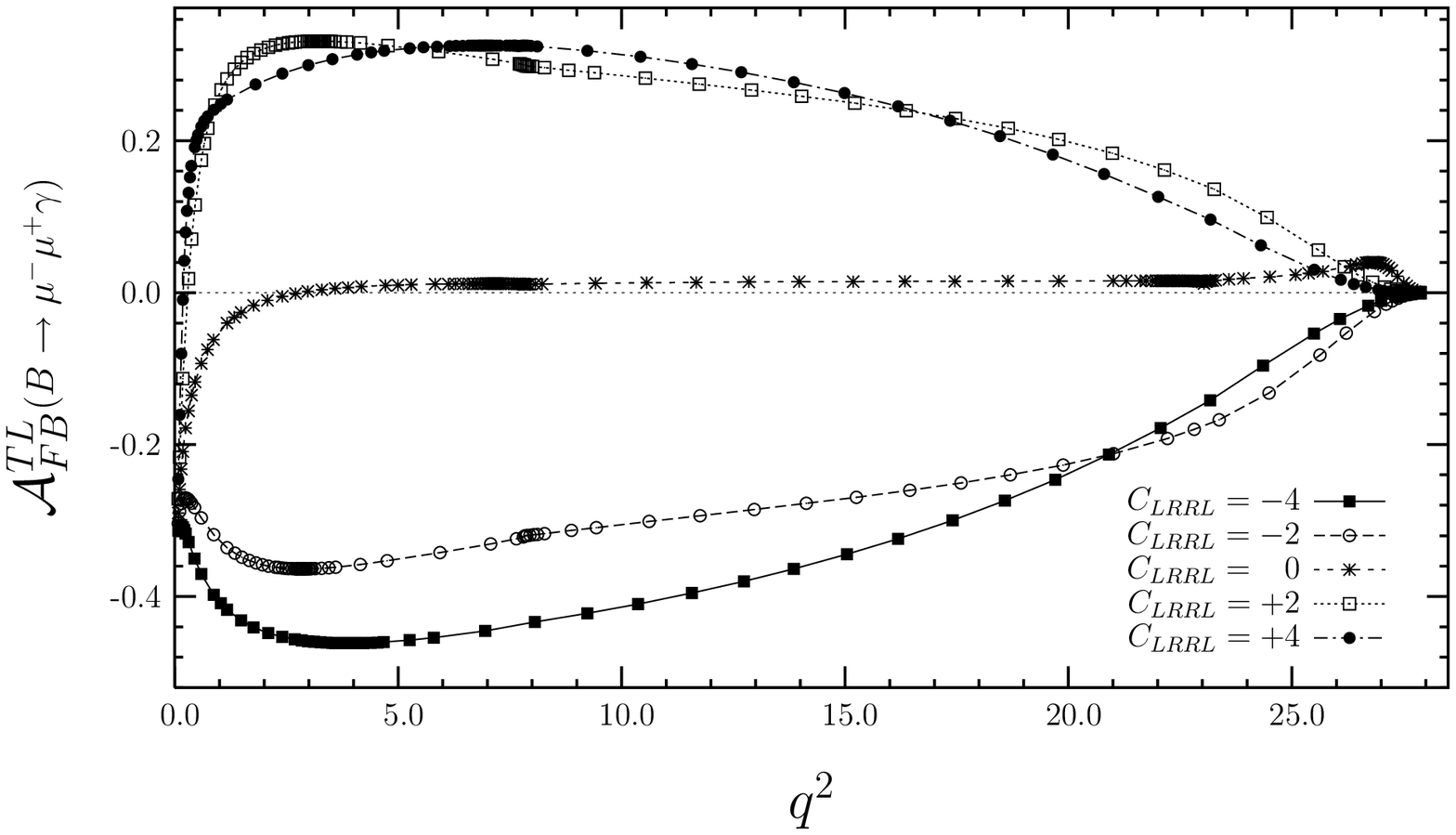}
\vskip 7.8 cm
\caption{}
\end{figure}

\begin{figure}
\vskip 2.5 cm
    \includegraphics{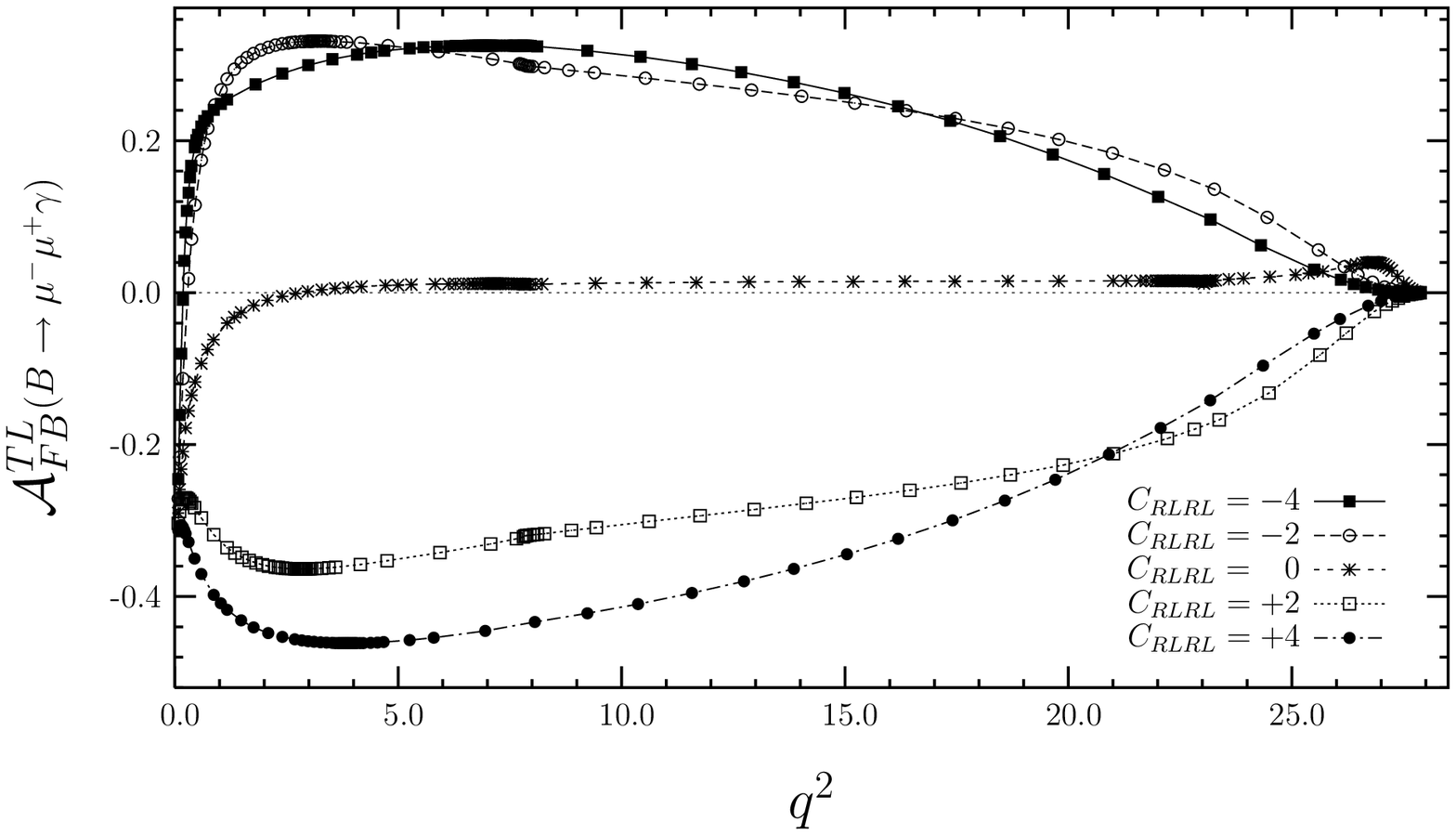}
\vskip 7.8 cm
\caption{}
\end{figure}

\begin{figure}
\vskip 2.5 cm
    \includegraphics{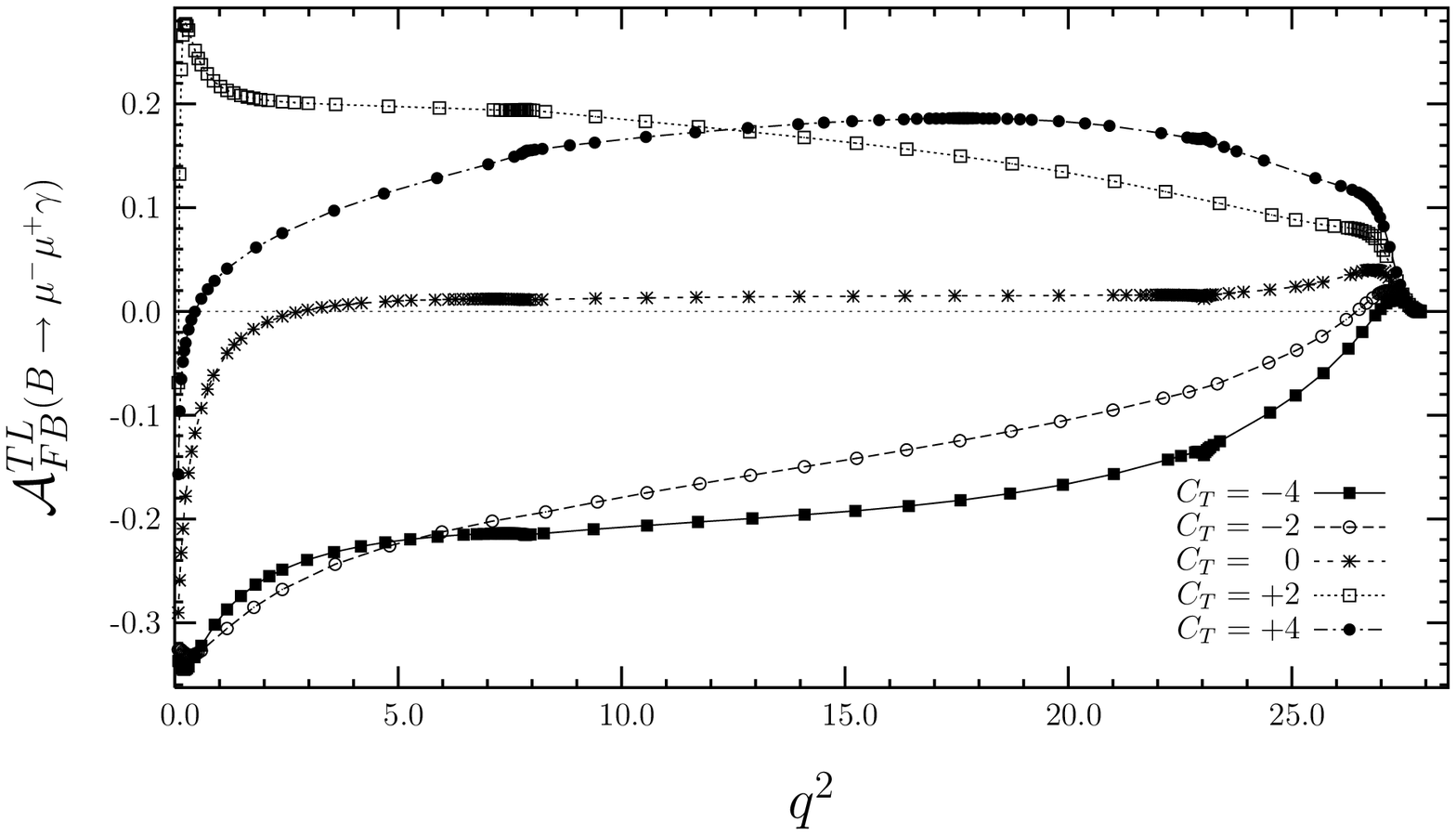}
\vskip 7.8 cm
\caption{}
\end{figure}

\begin{figure}
\vskip 2.5 cm
    \includegraphics{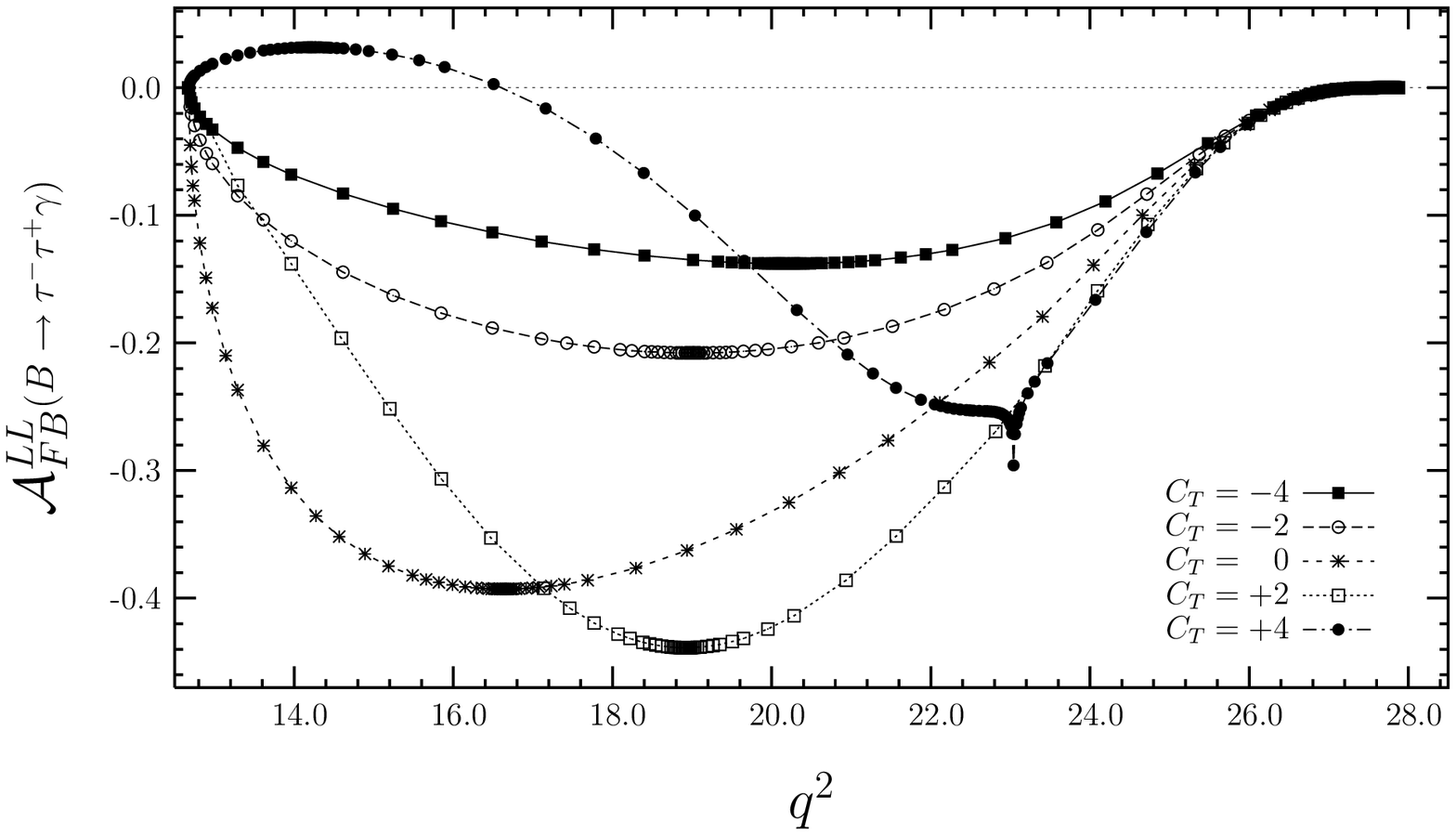}
\vskip 7.8 cm
\caption{}
\end{figure}

\begin{figure}
\vskip 1.5 cm
    \includegraphics{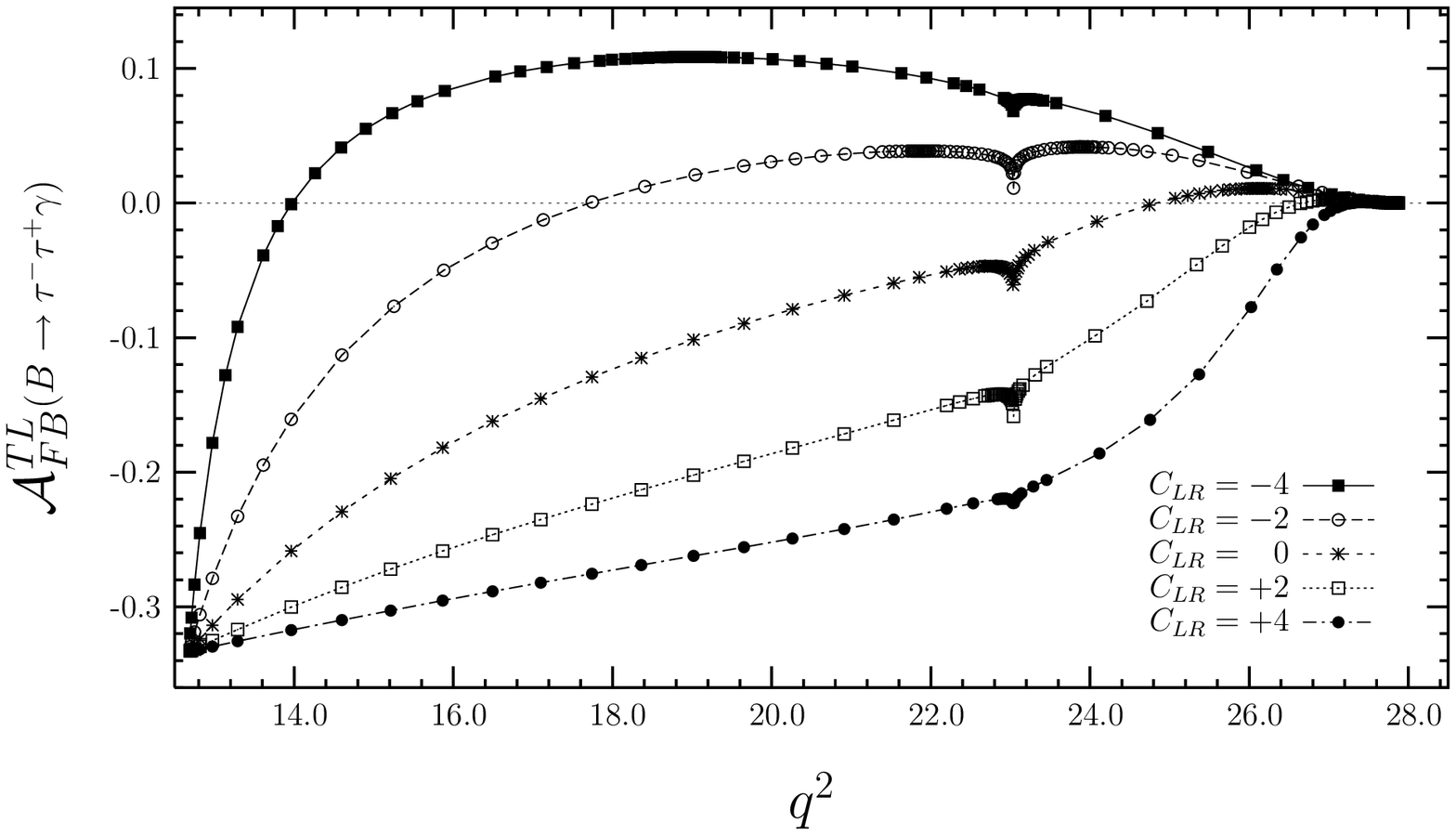}
\vskip 7.8cm
\caption{}
\end{figure}

\begin{figure}
\vskip 2.5 cm
    \includegraphics{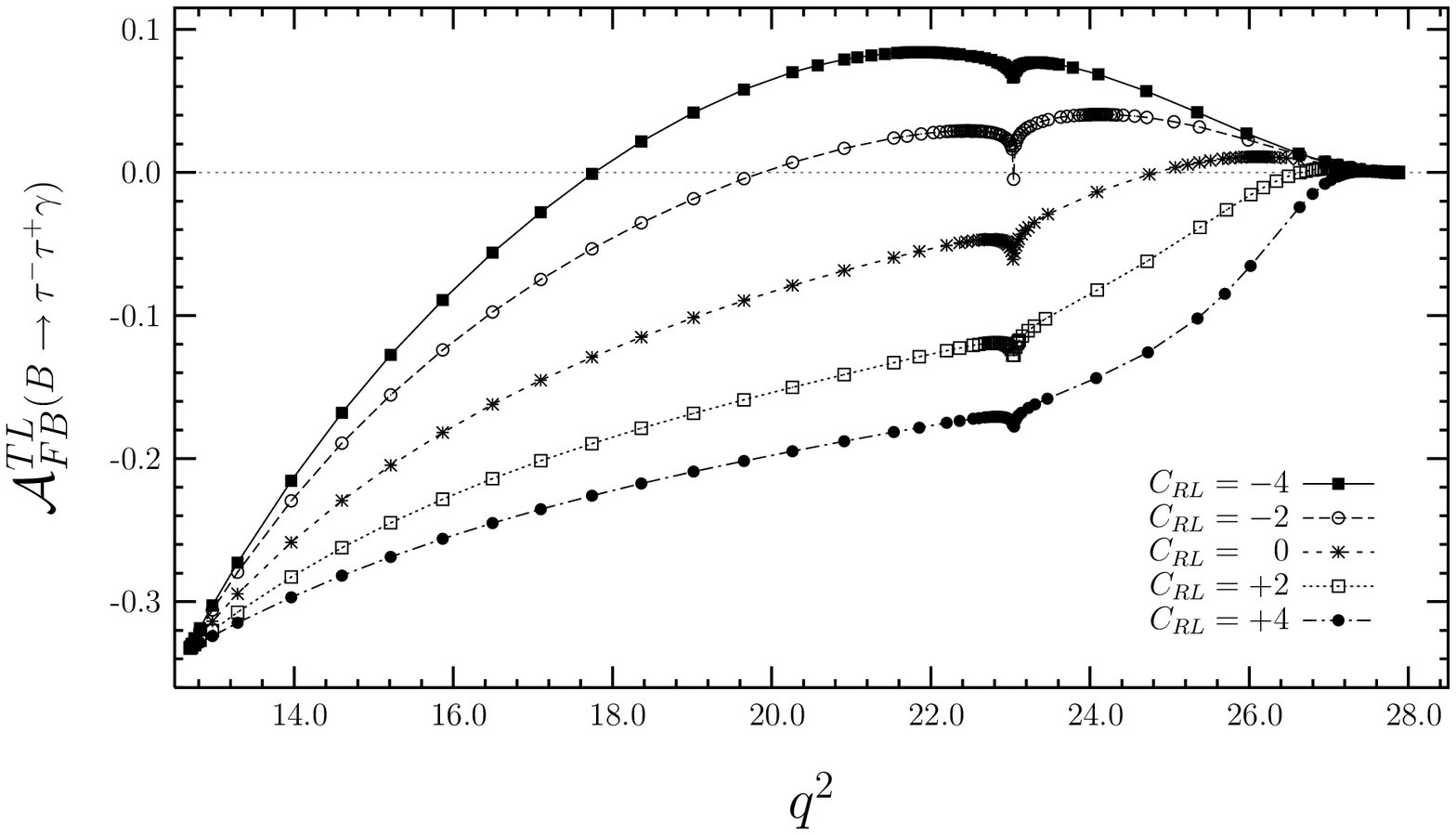}
\vskip 7.8 cm
\caption{}
\end{figure}

\begin{figure}
\vskip 1.5 cm
    \includegraphics{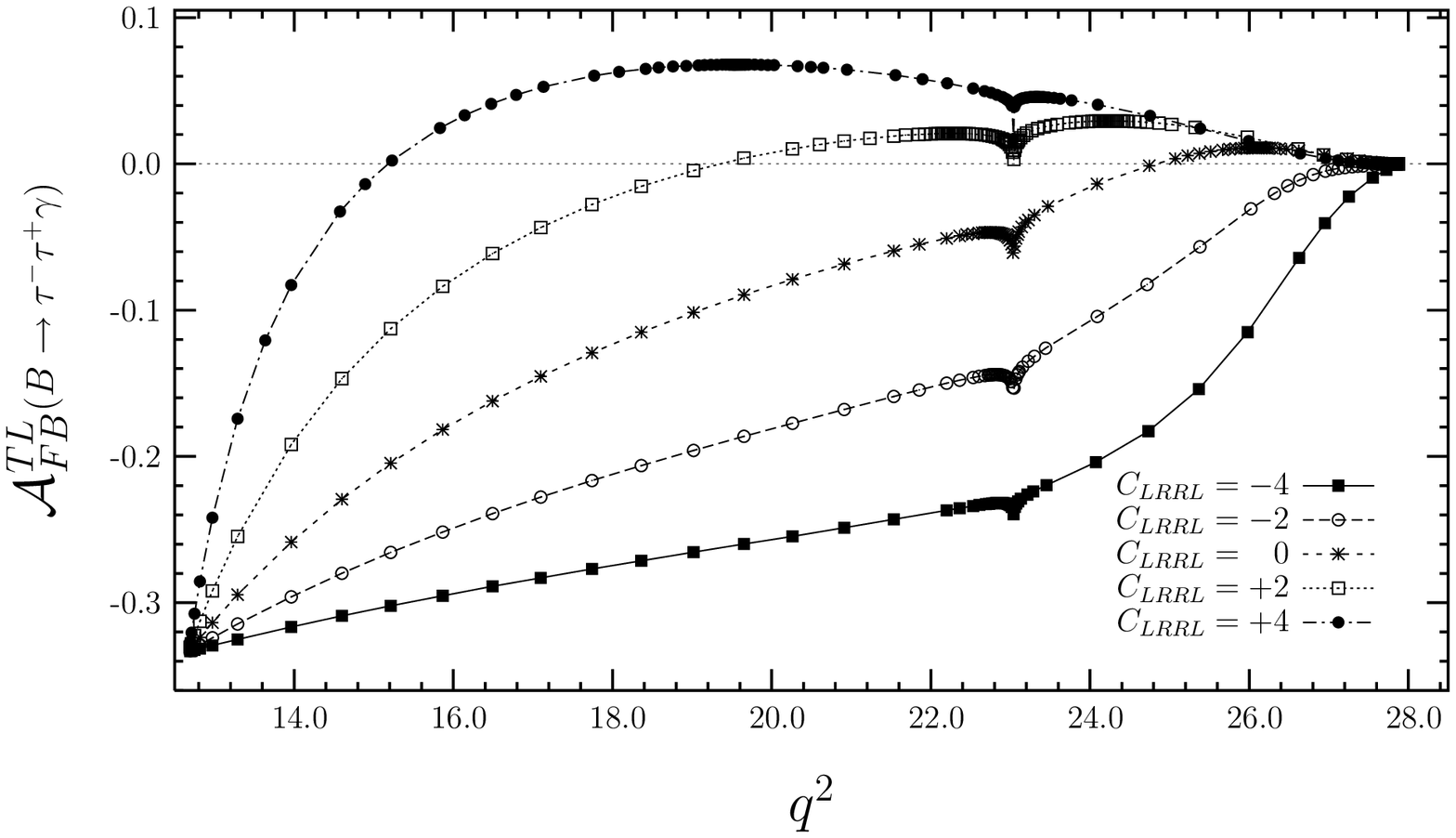}
\vskip 7.8cm
\caption{}
\end{figure}

\begin{figure}
\vskip 2.5 cm
    \includegraphics{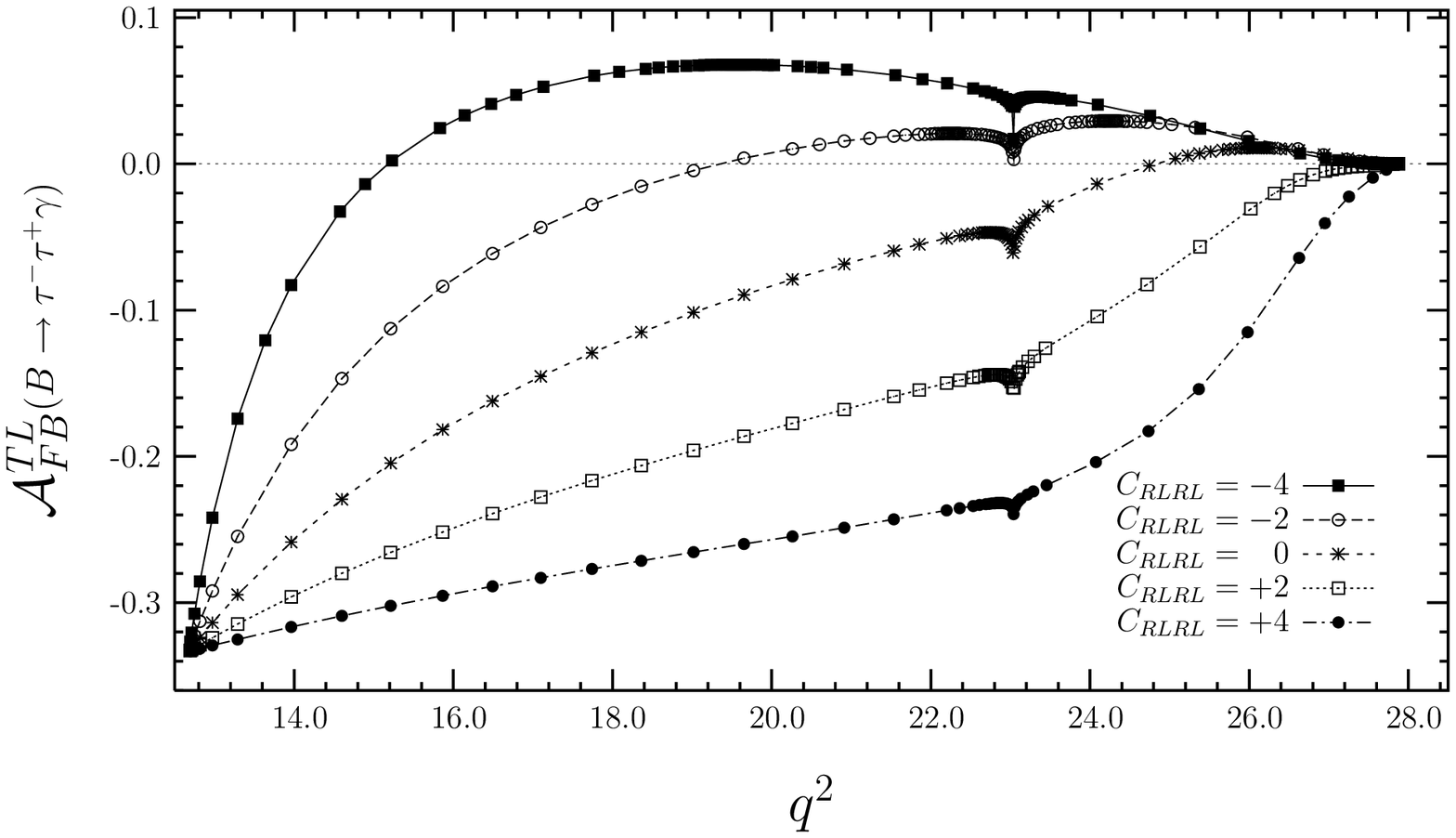}
\vskip 7.8 cm
\caption{}
\end{figure}

\begin{figure}
\vskip 2.5 cm
    \includegraphics{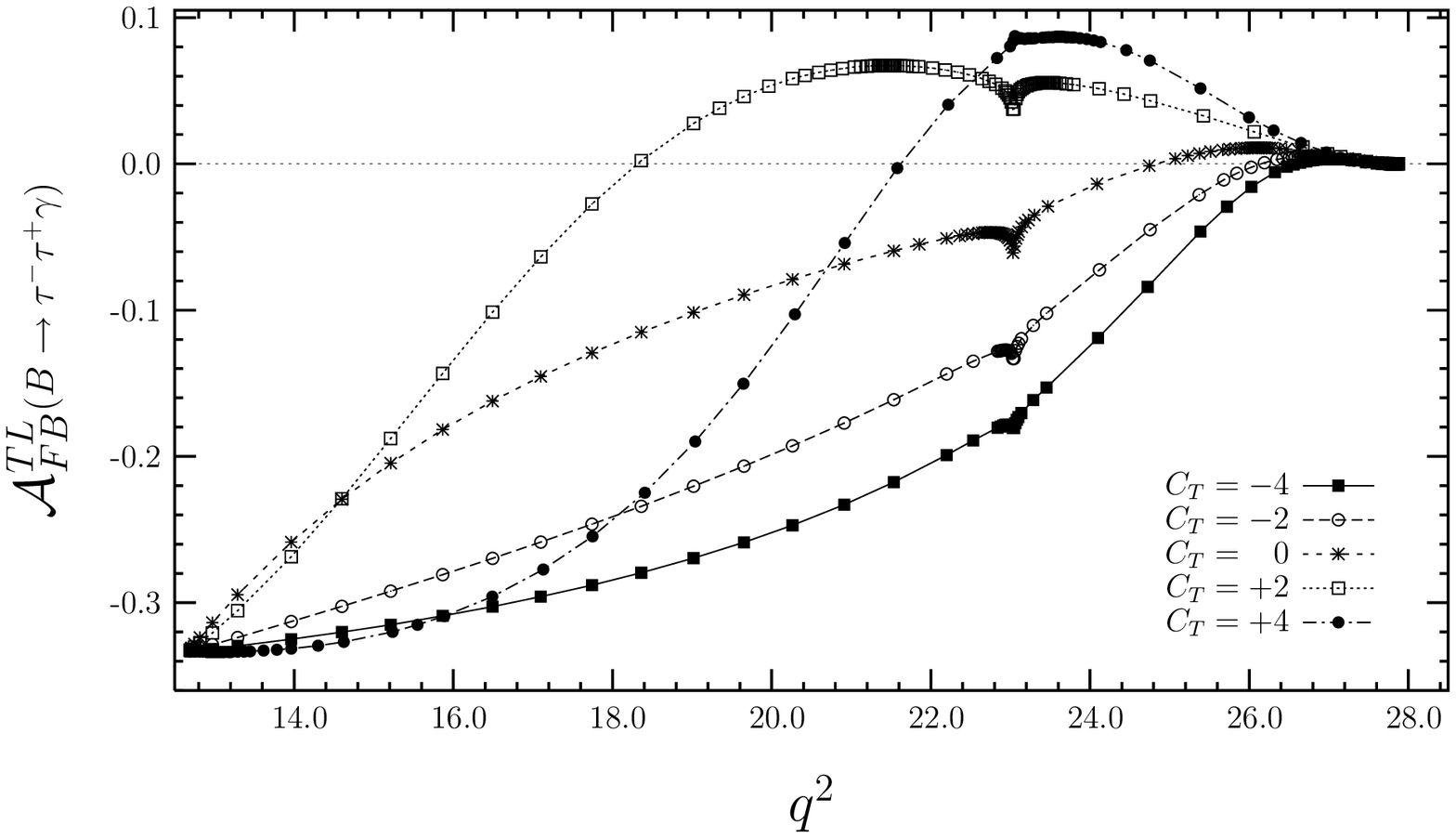}
\vskip 7.8 cm
\caption{}
\end{figure}

\begin{figure}
\vskip 1.5 cm
    \includegraphics{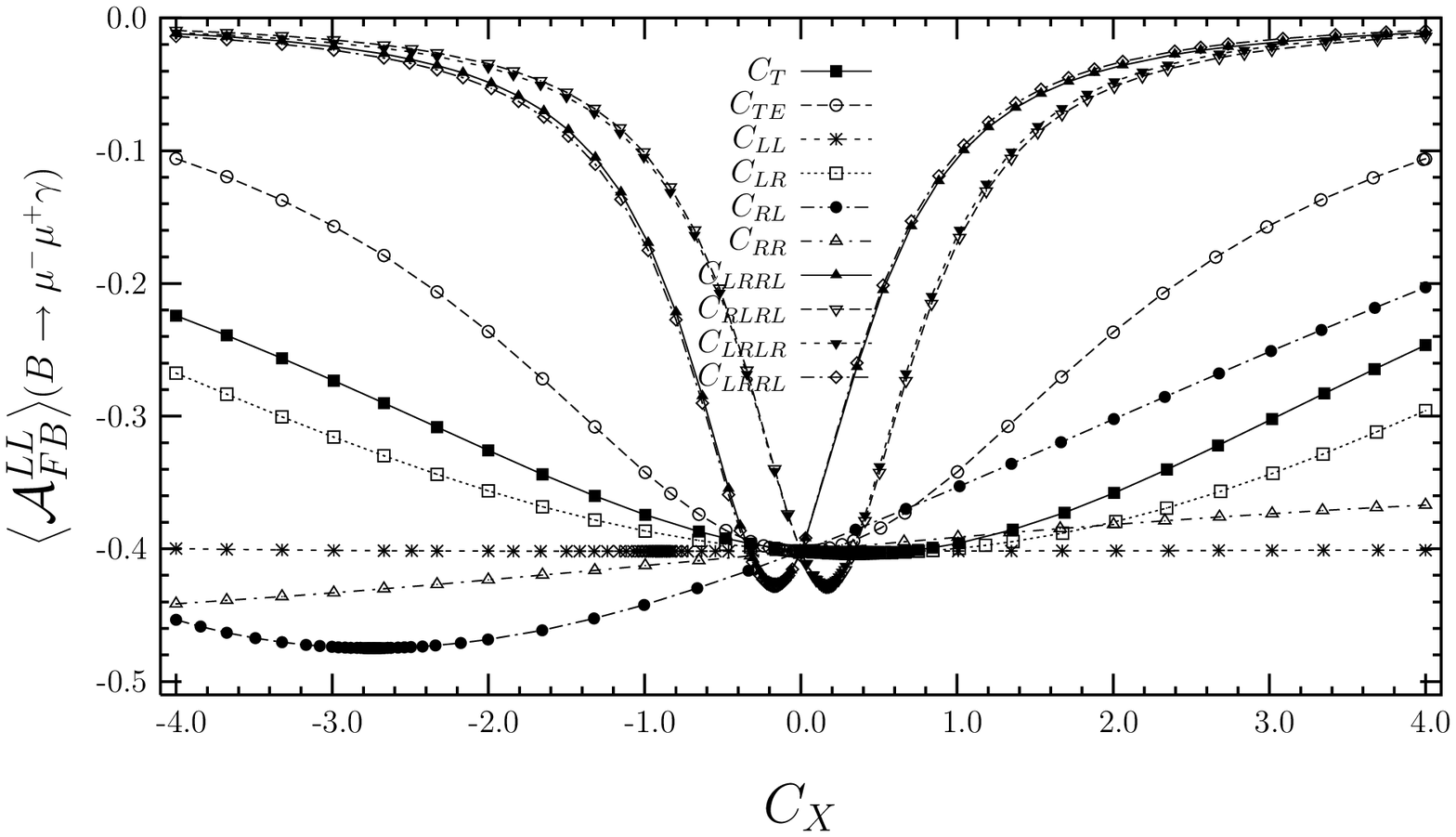}
\vskip 7.8cm
\caption{}
\end{figure}

\begin{figure}
\vskip 2.5 cm
    \includegraphics{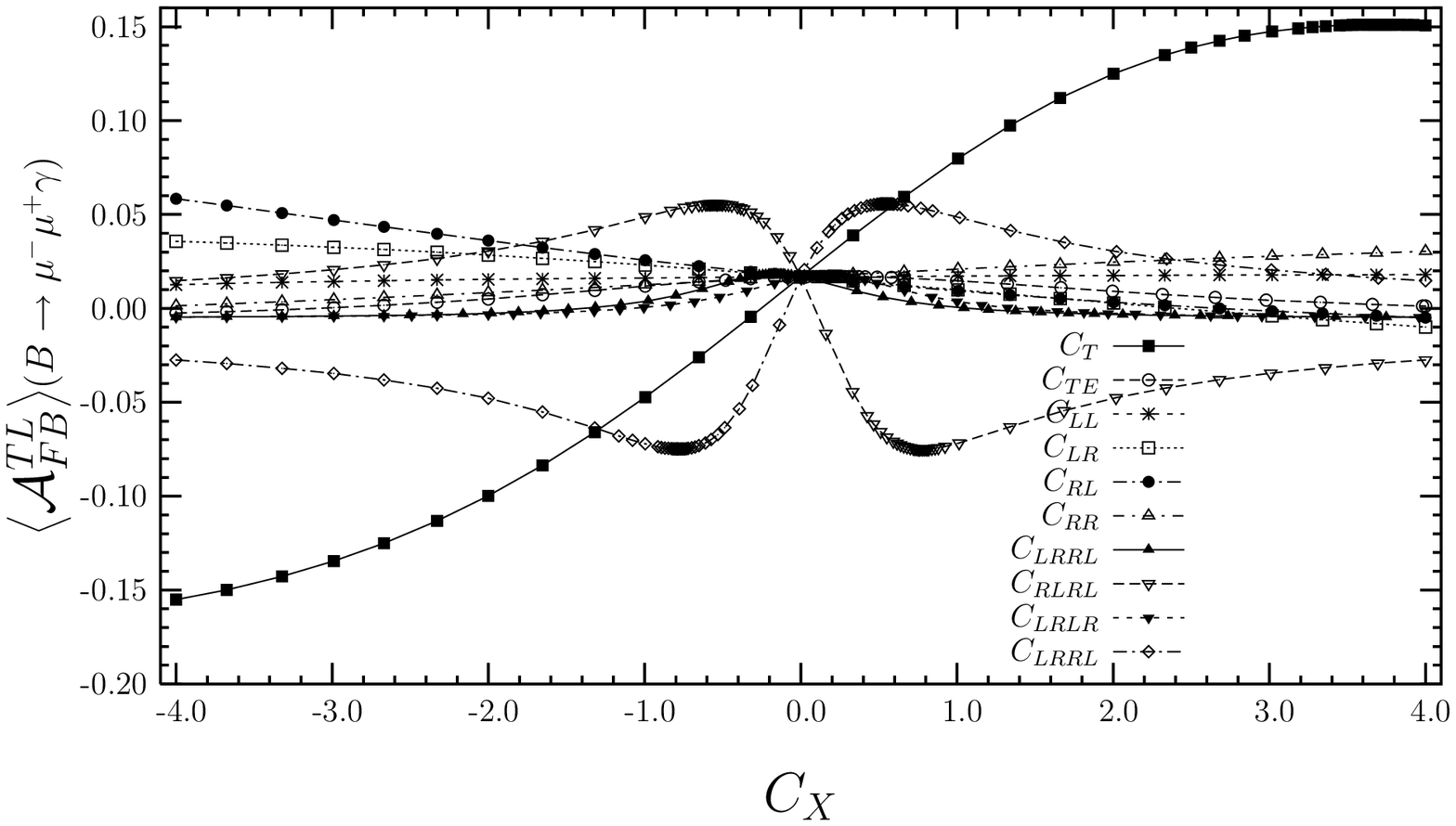}
\vskip 7.8 cm
\caption{}
\end{figure}

\begin{figure}
\vskip 2.5 cm
    \includegraphics{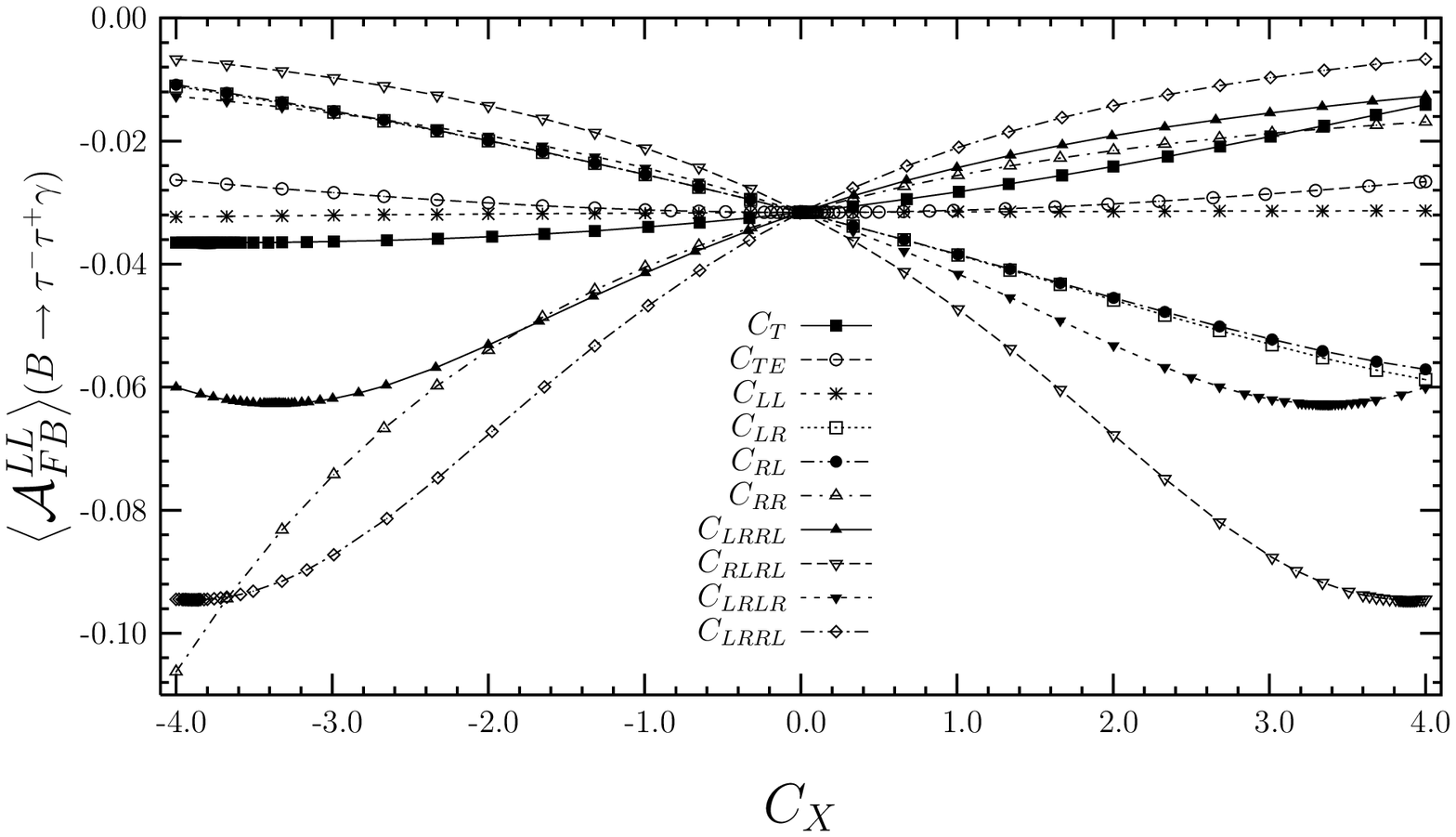}
\vskip 7.8 cm
\caption{}
\end{figure}

\begin{figure}
\vskip 2.5 cm
    \includegraphics{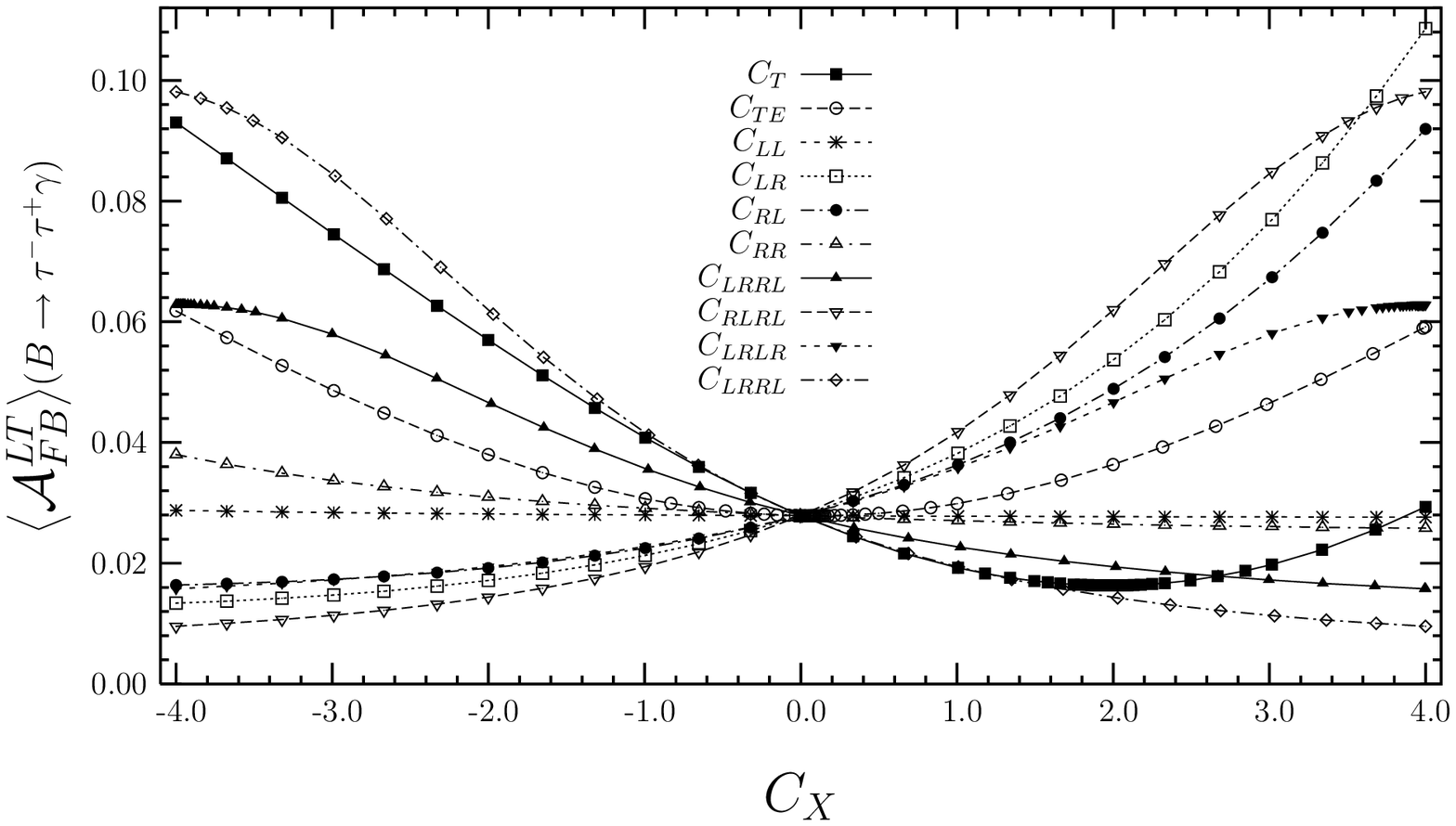}
\vskip 7.8 cm
\caption{}
\end{figure}

\begin{figure}
\vskip 2.5 cm
    \includegraphics{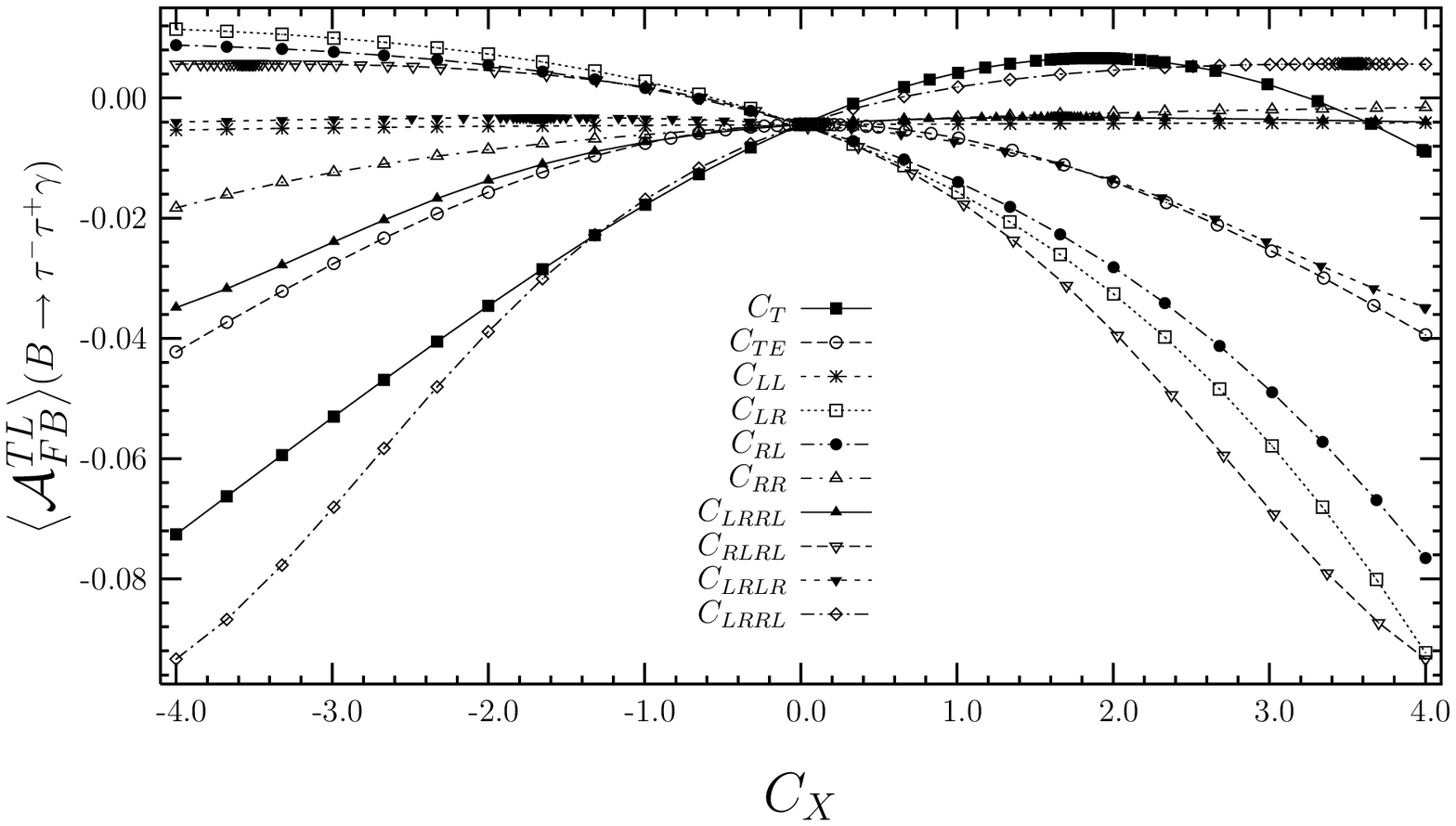}
\vskip 7.8 cm
\caption{}
\end{figure}

\end{document}